\documentclass[12pt]{spieman}  % 12pt font required by SPIE;
\usepackage{amsmath,amsfonts,amssymb}
\usepackage{graphicx}
\usepackage{setspace}
\usepackage{tocloft}
\usepackage{lineno}
 \usepackage[separate-uncertainty=true, multi-part-units=single]{siunitx}
\usepackage[colorlinks=true, allcolors=blue]{hyperref}
\usepackage{mathrsfs}
\usepackage{abbreviations}
\usepackage{xcolor} % Required for color support
\usepackage{makecell}
\usepackage{hyperref}
\usepackage{orcidlink}

\newcommand{\Rshunt}{R_\mathrm{sh}}
\newcommand{\Rn}{R_\mathrm{n}}
\newcommand{\Ib}{I_\mathrm{b}}
\newcommand{\Psat}{P_\mathrm{sat}}
\newcommand{\Popt}{P_{\gamma}}
\newcommand{\Pph}{P_{\phi}}
\newcommand{\Pbl}{P^\mathrm{bl}_{\gamma}}
\newcommand{\Pdd}{P^\mathrm{dd}_{\gamma}}
\newcommand{\Pod}{P^\mathrm{od}_{\gamma}}
\newcommand{\Pinband}{P^\mathrm{*}_{\gamma}}
\newcommand{\Tc}{T_\mathrm{c}}

\newcommand{\mK}{\,\mathrm{mK}}

\newcommand{\Tb}{T_\mathrm{b}}
\newcommand{\Tcmb}{T_\mathrm{cmb}}
\newcommand{\Trj}{T_\mathrm{RJ}}
\newcommand{\Lg}{\mathscr{L}_I}
\newcommand{\Ts}{T_\mathrm{s}}
\newcommand{\Ta}{T_\mathrm{a}}

\newcommand{\nlpf}{\eta_\mathrm{LPF}}
\newcommand{\fc}{\nu_\mathrm{c}}
\newcommand{\fcmb}{\nu_\mathrm{c}^\mathrm{cmb}}
\newcommand{\fdust}{\nu_\mathrm{c}^\mathrm{dust}}
\newcommand{\fsync}{\nu_\mathrm{c}^\mathrm{sync}}
\newcommand{\frj}{\nu_\mathrm{c}^\mathrm{RJ}}
\newcommand{\fbl}{\nu_\mathrm{c}^\mathrm{bl}}
\newcommand{\kb}{k_\mathrm{B}}
\newcommand{\nep}{\mathrm{NEP}}
\newcommand{\nepsq}{\mathrm{NEP}_\mathrm{sq}}
\newcommand{\nepph}{\mathrm{NEP}_{\phi}}
\newcommand{\nepopt}{\mathrm{NEP}_{\gamma}}
\newcommand{\nepbl}{\mathrm{NEP}_\mathrm{bl}}

\newcommand{\nepexc}{\mathrm{NEP}_\mathrm{exc}}
\newcommand{\flink}{F_\mathrm{link}}
\newcommand{\net}{\mathrm{NET}}
\newcommand{\nequ}{\mathrm{NEQ/U}}
\newcommand{\aWrtS}{\SI{}{\atto\watt\sqrt{\second}}}
\newcommand{\uKrtS}{\SI{}{\micro\kelvin\sqrt{\second}}}
\newcommand{\taur}{\tau_\mathrm{reion}}

\title{Detectors for CLASS-W2: The second 90 GHz telescope of the Cosmology Large Angular Scale Surveyor \vfill 
\vspace{2cm} % Adjust this spacing to fit your layout
\normalfont\large\textmd{Submitted to: \textit{Journal of Astronomical Telescopes, Instruments, and Systems, June 2026; SPIE Astronomical Telescopes + Instrumentation 2026
conference proceedings paper number 14156-55.}}}

\author[a]{John W. Appel\orcidlink{0000-0002-8412-630X}}
\author[b]{Kyuyoung Bae\orcidlink{0000-0002-3376-8660}}
\author[a]{Charles L. Bennett\orcidlink{0000-0001-8839-7206}}
\author[a]{Michael~K. Brewer}
\author[d]{Sarah Marie Bruno\orcidlink{0000-0003-2682-7498}}
\author[a]{Carol Yan Yan Chan\orcidlink{0000-0001-8144-556X}}
\author[a]{Joseph~Cleary\orcidlink{0000-0002-7271-0525}}
\author[g]{Sumit Dahal\orcidlink{0000-0002-1708-5464}}
\author[a]{Jullianna Denes~Couto\orcidlink{0000-0002-0552-3754}}
\author[f]{Kevin~L. Denis\orcidlink{0000-0002-3592-5703}}
\author[b]{Shannon M. Duff\orcidlink{0000-0002-9693-4478}}
\author[a]{Joseph~R. Eimer\orcidlink{0000-0001-6976-180X}}
\author[f]{Thomas Essinger-Hileman\orcidlink{0000-0002-4782-3851}}
\author[a]{Naina Gupta}
\author[b]{Johannes Hubmayr\orcidlink{0000-0002-2781-9302}}
\author[b,c]{Gregory Jaehnig\orcidlink{0000-0001-8697-0064}}
\author[a]{John Karakla}
\author[b,c]{Matthew Koc\orcidlink{0000-0003-1401-8415}}
\author[b]{Jeff Van Lanen\orcidlink{0009-0009-8245-7429}}
\author[h]{Yunyang Li\orcidlink{0000-0002-4820-1122}}
\author[b]{Michael J. Link\orcidlink{0000-0003-2381-1378}}
\author[b]{Tammy Lucas\orcidlink{0000-0001-7694-1999}}
\author[a]{Tobias Marriage\orcidlink{0000-0003-4496-6520}}
\author[a]{Carolina Morales Perez\orcidlink{0000-0002-1371-5334}}
\author[e]{Matthew A. Petroff\orcidlink{0000-0002-4436-4215}}
\author[a]{Caleigh Ryan\orcidlink{0009-0001-1748-7877}}
\author[i]{Rui Shi\orcidlink{0000-0001-7458-6946}}
\author[a]{Deniz~A.~N. Valle\orcidlink{0000-0003-3487-2811}}
\author[f]{Edward~J. Wollack\orcidlink{0000-0002-7567-4451}}

\affil[a]{The William H. Miller III Department of Physics and Astronomy, Johns Hopkins University, 3400 N. Charles Street, Baltimore, 21218, MD, USA}

\affil[b]{Quantum Sensors Division, NIST, 325 Broadway, Boulder, 80305, CO, USA}
\affil[c]{Department of Physics, University of Colorado, 390 UCB, Boulder, 80309, CO, USA}

\affil[d]{Department of Physics, Villanova University,  Villanova, PA 19085, USA}

\affil[e]{Center for Astrophysics, Harvard \& Smithsonian, 60 Garden Street, Cambridge, MA 02138, USA}

\affil[f]{NASA Goddard Space Flight Center, 8800 Greenbelt Road, Greenbelt, MD 20771, USA}
\affil[g]{Johns Hopkins University Applied Physics Laboratory, 11100 Johns Hopkins Road, Laurel, MD 20723}

\affil[h]{Kavli Institute for Cosmological Physics, University of Chicago, 5640 South Ellis Avenue, Chicago, IL 60637, USA}

\affil[i]{Department of Physics, Stanford University, Stanford, CA 94305, USA Kavli Institute for Particle Astrophysics and Cosmology, Stanford, CA 94305, USA}

\cftpagenumbersoff{figure}
\cftpagenumbersoff{table} 
\begin{document}

\maketitle
\begin{abstract}
The Cosmology Large Angular Scale Surveyor (CLASS) is measuring the Cosmic Microwave Background (CMB) polarization anisotropy on the largest angular scales ($>\SI{1}{\deg}$) to probe the epochs of inflation and reionization. To enhance the CMB mapping speed, we have built, tested, and commissioned in August 2025 a second \SI{90}{\giga\hertz} receiver (CLASS-W2) with a detector focal plane composed of feedhorn-coupled Transition Edge Sensor (TES) bolometers fabricated at NIST-Boulder. The focal plane consists of four modules, each containing 37 feedhorns coupling orthogonal polarizations onto two TES bolometers, for a total of 296 optically sensitive detectors. Laboratory tests show highly uniform TES properties with an array average critical temperature of \SI{184\pm3}{\milli\kelvin}, a thermal conductance of \SI{460\pm47}{\pico\watt\per\kelvin}, and a normal resistance of \SI{7.8\pm0.3}{\milli\ohm}.  The detector array has an average band center frequency of \SI{95.2}{\giga\hertz} with \SI{28.3}{\giga\hertz} bandwidth, and achieves a detector yield of 94\%. On-sky measurements indicate a mean detector optical load of \SI{3.3}{\pico\watt}, corresponding to an antenna temperature of $\sim\SI{23}{\kelvin}$.  The array’s average beam solid angle is \SI{124}{\micro\steradian}, with a full width at half maximum of $\SI{0.592}{\deg}$, and the end-to-end average optical efficiency is 0.37. We find that high-frequency “blue-leak’’ radiation couples directly to the TES bolometer islands; adding a metal-mesh low-pass filter with cutoff frequency of \SI{157}{\giga\hertz} in front of the focal plane suppresses the “blue-leak’’ power by \SI{0.9}{\pico\watt}. The four-module array achieves a noise-equivalent temperature of $\net= 16~\uKrtS$. Adding this array has boosted the CLASS \SI{90}{\giga\hertz}  mapping speed by 41\%. 

%Three more detector modules are now in production and will complete the seven-module CLASS-W2 detector focal plane, scheduled to be deployed in 2027.

 \end{abstract}


% Include email contact information for corresponding author
{\noindent \footnotesize\textbf{*}John Appel,  \linkable{jappel3@jhu.edu} }

\begin{spacing}{1}   % use 2 for double spacing for rest of manuscript

\section{Introduction}\label{sec1}

Among the six base parameters of the standard $\Lambda$CDM~\cite{bennett13,planck18I} model of cosmology that describes the large-scale structure, contents, and evolution of the Universe, the optical depth to reionization, $\taur$\footnote{We denote the optical depth to reionization as $\taur$ to distinguish it from the detector thermal time constant $\tau$.}, remains the least constrained. Space-based measurements of the large-scale ($\ell < 30$) Cosmic Microwave Background (CMB) polarization signal disagree, giving $\taur = 0.089\pm0.014$~\cite{Hinshaw13}\footnote{When the WMAP polarization data are cleaned using the Planck \SI{353}{\giga\hertz}  map as a polarized dust template, the inferred optical depth to reionization is $\taur \simeq 0.075 \pm 0.013$\cite{planckXLVII}.} and $\taur = 0.057\pm0.006$~\cite{pagano19,tristram24}; differences that propagate into the inferred amplitude of primordial fluctuations and the sum of neutrino masses. If the low-$\ell$ polarization data are set aside, a $\Lambda$CDM fit that combines DESI BAO, high-$\ell$ CMB spectra, and CMB lensing instead prefers $\taur = 0.09\pm0.0012$~\cite{sailer26}.

The Cosmology Large Angular Scale Surveyor is the first ground-based CMB experiment to constrain $\taur$. Cross-correlating the first CLASS \SI{90}{\giga\hertz} map with Planck data results in $\taur = 0.055\pm0.019$ \cite{li25}.
In contrast to WMAP and Planck, which have finished collecting data, CLASS remains an active experiment, continually observing the CMB, upgrading its instruments and scan strategy, refining its analysis pipeline, and adding new telescopes to further increase its mapping speed.
In this report, we present the design and characterize the performance of the detector focal plane for the second CLASS \SI{90}{\giga\hertz} receiver, referred to as CLASS-W2. 

Section~\ref{sec:telescope} describes the CLASS-W2 telescope, the detector focal plane, the components and assembly of the detector modules, and the detector microwave design that couples the CMB signal to the Transition Edge Sensor (TES) bolometer.
Section~\ref{sec:tes_param} presents the TES detector characterization based primarily on I--V data acquired in lab with no optical power illuminating the detectors.
Section~\ref{sec:tes_param_opt} discusses the optical characterization of the detectors, including bandpass measurements, Jupiter beam maps, detector optical efficiency measurements, out-of-band power constraints and mitigation, on-sky noise performance, and finally the array mapping sensitivity. 
We end by summarizing the results in Section~\ref{sec:conclusions}.

\section{CLASS-W2 Telescope}\label{sec:telescope}

The CLASS telescopes observe from a high-altitude (\SI{5200}{\meter}) site in the Atacama Desert of Chile, covering over 75\% of the sky by continuously scanning \SI{720}{\degree} in azimuth at constant elevation (\SI{45}{\degree}) and boresight angles. The boresight angle is changed every day by \SI{15}{\degree} in between the range of \qtylist[list-units = single]{-45; 45}{\degree}.
The sky polarization signal is modulated at $\sim\SI{10}{\hertz}$ with a front-end polarization modulator. CLASS has successfully deployed and operated both Variable-delay Polarization Modulators (VPM)\cite{chus12vpm}, and rotating Reflective Half Wave Plates (RHWP)\cite{shi24-spie}. The VPM modulates $U$ and $V$ Stokes parameters providing sensitivity to circular polarization, while the RHWP is optimized for linear polarization, modulating Stokes parameters $U$ and $Q$ at four times the rotation frequency ($4f$). The RHWP signal band at $4f$ is separate from the RHWP synchronous emission/reflection modulated at $2f$, adding another layer of systematic control. Additionally, the RHWP modulates all polarization angles on the sky at each telescope boresight.

The first \SI{90}{\giga\hertz} telescope, CLASS-W1, was installed in 2018 and received a major detector upgrade in 2022. To improve sensitivity at \SI{90}{\giga\hertz}, the CLASS frequency band most sensitive to the CMB, we developed and deployed a second \SI{90}{\giga\hertz} telescope, CLASS-W2. First light was observed in August 2025 coupled to a VPM. In February 2026, a free-space metal-mesh low-pass filter with a cutoff at \SI{157}{\giga\hertz}\cite{ade06} was installed in front of the detector focal plane on the \SI{50}{\milli\kelvin} stage to eliminate a low level of `blue leak' optical power detected during the initial deployment. In May 2026, after making pioneering new measurements of the circular polarization of foregrounds\cite{petroff20}, and setting the best current upper limits on circular CMB polarization\cite{padilla20}, we turned our full attention to linear polarization, the VPM was replaced with a RHWP to improve sensitivity to linear polarization.

\subsection{Telescope Optics}

The CLASS-W2 telescope is similar in design to the CLASS-W1\cite{datta24,nunez25} telescope, but observes from the second CLASS mount. Deploying two \SI{90}{\giga\hertz} telescopes observing from two different mounts simultaneously allows or better characterization and mitigation of systematics. 
Briefly, the telescope sits inside a protective cage with a large forebaffle extending from the cage opening to reduce sidelobe pickup. The first optical element is a VPM or RHWP that modulates the polarization of the sky signal. Both modulators are reflective, with no dielectric materials. Two reflective mirrors then couple the modulator output to the cryostat window. Inside the cryostat, two cryogenically cooled high-density polyethylene (HDPE) lenses focus the light onto the detector focal plane. Details of the optical design of the telescope can be found in Ref.~\citenum{datta24}. Infrared filters reduce thermal loading on the \SI{60}{\kelvin}, \SI{4}{\kelvin}, \SI{1}{\kelvin}, and \SI{50}{\milli\kelvin} stages \cite{iuliano2018spie}. CLASS uses multi-layer IR-absorptive foam filters behind the window, polytetrafluoroethylene (PTFE) absorptive filters at \SI{60}{\kelvin} and \SI{4}{\kelvin}, and a Nylon filter at \SI{1}{\kelvin}.  A Bluefors\footnote{\url{https://bluefors.com/}} dilution refrigerator cools the detector focal plane to below \SI{50}{\milli\kelvin} during observations. Cylindrical magnetic shields at the  \SI{4}{\kelvin} and \SI{50}{\milli\kelvin} stages surround the focal plane, reducing the susceptibility of the TES bolometers and  superconducting quantum interference device (SQUID) readout to Earth’s magnetic field and to its modulation from azimuth scanning.

\subsection{Detector focal plane}

\begin{figure} 
   \begin{center}
   \label{fig:fp_pic} 
   \begin{tabular}{c} 
   \includegraphics[height=6.7cm, trim={0cm 0cm 0cm 0cm},clip]{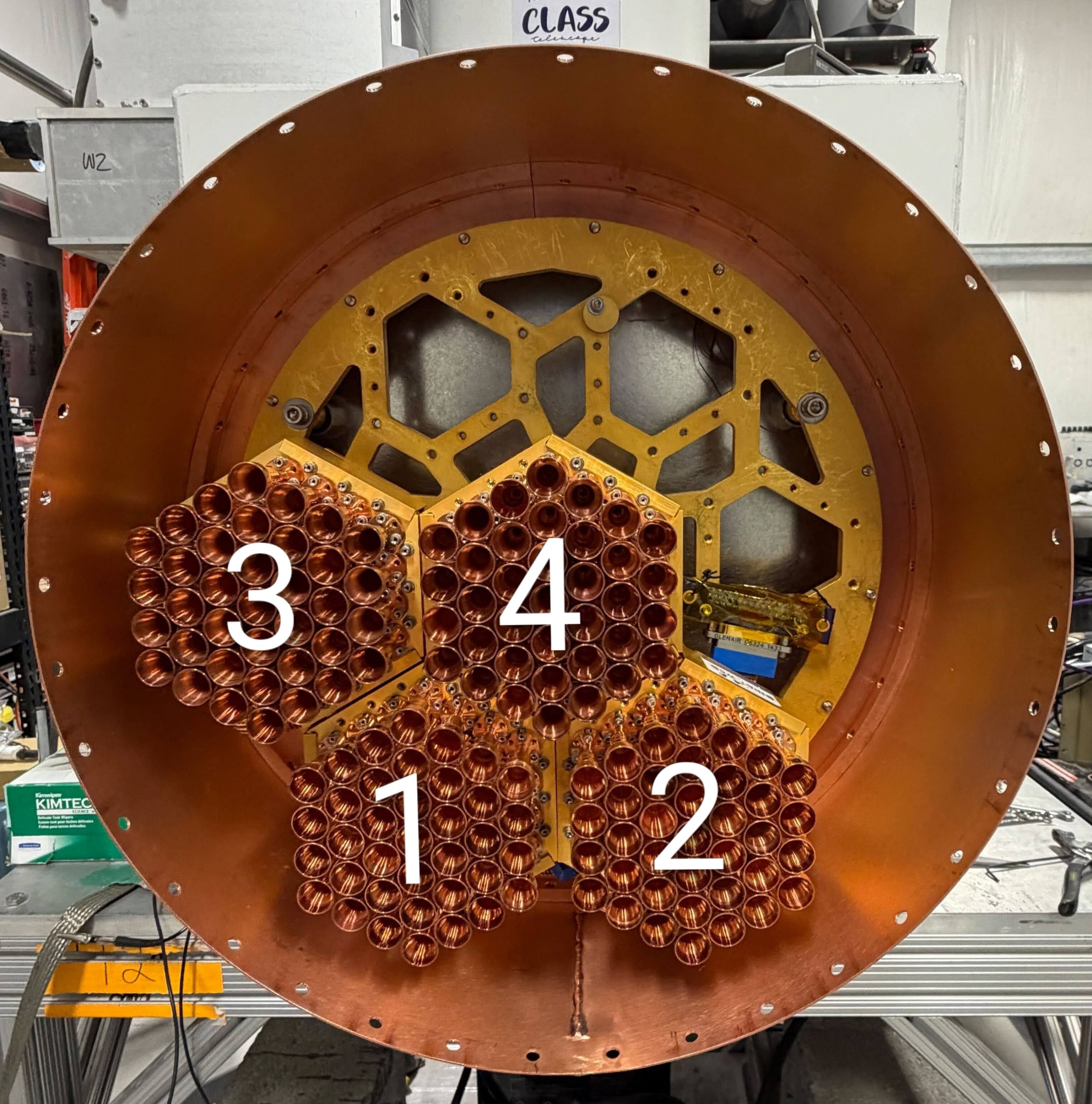}
    \includegraphics[height=6.7cm, trim={0cm 0cm 0cm 0cm},clip]{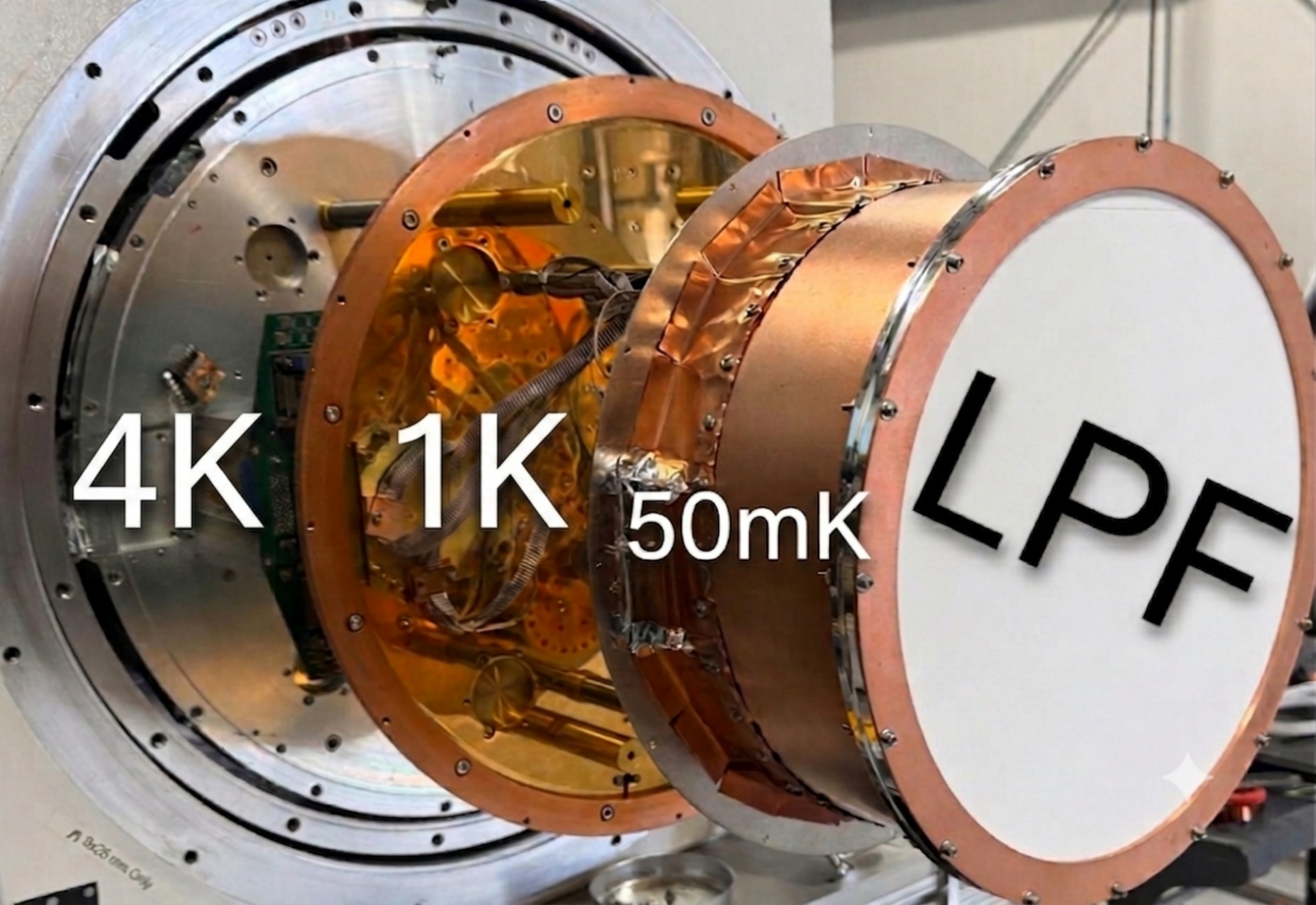}
    \end{tabular}
    \end{center}
   \caption[CLASS-W2 photographs]
   {
   \textbf{Left:}  photo of the four \SI{90}{\giga\hertz} detector modules mounted on the CLASS-W2 \SI{50}{\milli\kelvin} stage. Module labels 1, 2, 3, and 4 indicate the modules’ locations on the CLASS-W2 focal plane and correspond to wafer fabrication numbers 3, 5, 1, and 2, respectively. Wafer 4 was not used in this deployment run. We are currently fabricating three new detector modules that will be installed in 2027 to the three empty slots at the top-right of the focal plane. \textbf{Right:}  photo of Cardiff LPF installed in 2026 in front of the detector modules. The \SI{50}{\milli\kelvin}, \SI{1}{\kelvin}, and \SI{4}{\kelvin} temperature stages are labeled at the cryostat baseplates. }
   \end{figure}

The light reaching the focal plane couples into precision-machined copper feedhorns\cite{zeng10spie}.
Thirty-seven feedhorns are mounted on one side of a module's SiAl composite baseplate~\cite{ali18spie,ali22}. Each feedhorn feeds a circular waveguide that delivers the polarization signals to the detector’s superconducting planar ortho-mode transducer (OMT) antenna probes. A full CLASS \SI{90}{\giga\hertz} focal plane comprises seven modules; the initial CLASS-W2 focal plane (Figure~\ref{fig:fp_pic}) was deployed in 2025 and includes four modules. A second detector fabrication run is underway to complete the seven-module focal plane. The seven modules mount onto a copper-web that is well heatsunk to the \SI{50}{\milli\kelvin} cryostat stage. A Ruthenium oxide (Rox) thermometer mounted on the copper-web tracks the detector bath temperature ($\Tb$), while a copper-can bolted to the edge of the web holds a free-space metal-mesh low-pass filter (LPF)\cite{ade06} as shown in the right panel of Figure~\ref{fig:fp_pic}. The LPF is clamped and heatsunk to a copper ring with a spiral spring gasket\cite{iuliano20thesis}. The LPF together with the copper can  assembly were deployed in 2026 to form a light-tight enclosure that only allows radiation below \SI{157}{\giga\hertz} to reach the focal plane. Before the LPF installation, out-of-band power contributed significantly to detector loading, as described in Section~\ref{sec:optical_loading}.

%the radiation reaching the focal plane peaked near $\sim\SI{300}{\giga\hertz}$ with higher frequencies more strongly absorbed by the \SI{1}{\kelvin} Nylon filter.    
 
The NIST-fabricated detector wafers are integrated with a backshort-cap, backshort, waveguide interface plate (WIP), and choke. The backshort includes moats around the waveguide that are filled with dark epoxy to absorb stray light. A similar NIST detector integrated wafer design is presented in \citenum{ward16spie}. Figure~\ref{fig:det_stack} in the appendix shows a cross-sectional view of a CLASS detector wafer. The detector wafers mount on the SiAl baseplate, on the opposite side to the feedhorns. The detector wafer is aligned and secured using the same pin-and-spring mechanical constraints as the previous CLASS-W1 telscope modules that used NASA-fabricated detector wafers \cite{dahal18spie, nunez22spie}. Over two hundred gold bonds are placed between the gold plated surfaces of the detector wafer and the gold plated SiAl baseplate. These gold bonds improve the thermal heat sinking of the detector wafer to the \SI{50}{\milli\kelvin} stage.
%increasing detector saturation power, detector speed, and decreasing phonon noise. 
The TES bolometers are connected with superconducting aluminum wire bonds to the shunt resistors and SQUID readout. The cryogenic circuit board and cables that carry the detector signal to the \SI{300}{\kelvin} Multi-Channel Electronic (MCE) readout are identical to those deployed in the CLASS-W1 telescope\cite{Battistelli08}.

\subsection{Microwave architecture and TES design}
The OMT waveguide probes are impedance matched to the on-chip co-planar waveguide (CPW) lines, which transition to low-impedance microstrip (MS) lines built on SiN dielectric.
The orthogonal polarization signals carried by the MS lines are combined in a hybrid-tee\cite{knochel90} where unwanted sum modes are terminated. The MS continues to the on-chip band-defining filters before being terminated on a lossy gold meander adjacent to the AlMn TES on the SiN bolometer island. A PdAu thermal ballast sets the heat capacity of the TES bolometer, which determines the stability and time constant of the detector response. A more detailed description of the detector microwave design is presented in Ref.~\citenum{jaehnig26}~(Figure~1), including a photo of the fabricated microwave detector components.

\section{TES detector parameters}\label{sec:tes_param}
The TES bolometer parameters are measured in-lab with the detectors mounted inside a dark \SI{1}{\kelvin} cold copper cavity, allowing us to characterize the detector under effectively zero optical loading ($\Popt = 0$). 
CLASS detectors are read by cryogenically cooled SQUID\cite{nist_tdm_mux13b} that are multiplexed in the time domain with an MCE warm readout system\cite{Battistelli08}. Each CLASS \SI{90}{\giga\hertz} module is addressed by 4 columns with 22 multiplexed rows. A full seven module CLASS \SI{90}{\giga\hertz} focal plane contains 28 columns and a total of 616 readout channels. 
Each TES bolometer is voltage biased by a small $\SI{250}{\micro\ohm}$ shunt resistor ($\Rshunt$) fabricated by NIST. The bias current ($\Ib$) is applied by low-noise MCE bias lines. The SQUID input is connected in series with the TES to measure changes in the TES current, which are logged by the MCE readout cards. For a circuit diagram of the TES voltage biased circuit see Ref.~\citenum{appel22}~(Figure~1).

The absolute TES current ($I$) is determined from I--V curve measurements. These measurements consist of the recorded change in TES current as the TES voltage $V$ (or equivalently the bias current $\Ib$) is linearly decreased from a high value down to zero; tracing out an I--V curve. The voltage sweep starts at a high voltage such that the TES is operating as a normal resistor with resistance $\Rn$. As we decrease the voltage the bias power dissipated on the TES bolometer island drops, and therefore the TES temperature also decreases until it reaches its superconducting critical temperature ($\Tc$). At $\Tc$ the TES becomes sensitive to small temperature fluctuations and operates under negative electro-thermal feedback on its superconducting transition. As the voltage continues to drop eventually the TES ``falls-off'' the transition and becomes superconducting. See an example of an I--V curves collected at multiple bath temperatures in Figure~\ref{fig:iv_G}. 

The $y$-intercept of a linear fit to the normal branch of the I--V curve, found at the high end of the voltage bias, provides the absolute current offset used to convert the measured change in TES current into an absolute current.
 With knowledge of both TES $I$ and $V$ we can compute the TES resistance $R$ and TES bias power $P$. In particular we define the saturation power $\Psat$, as $P$ measured at $R$ equal to 70\% of $\Rn$. $\Psat$ represents the maximum amount of power a bolometer can absorb while operating on the superconducting transition, or equivalently the amount of phonon power ($\Pph$) that flows from the TES volometer island to the bath when the TES is operating on-transition and there is zero optical loading ($\Popt = 0$) on the bolometer.

The saturation power extracted from I--V curves acquired at multiple bath temperatures ($\Tb$) are used in the following sections to determine the TES thermal conductivity parameters for each detector in the CLASS-W2 focal plane. From I--V curves we also determine the TES responsivity calibration ($S$) to convert measured TES current to power. The ratio of the responsivities calculated based on the slope of a single I--V curve versus multi-$\Tb$ I--V curves yields an estimate of the TES loop gain $\Lg$ and the logarithmic temperature sensitivity of the superconducting transition $\alpha = \frac{T}{R}\frac{dR}{dT}$.  Finally, on-sky time constants measured from the phase of the polarization modulation synchronous signal are compared to the expected TES time constants derived from dark multi-$\Tb$ I--V curve $\alpha$ estimates.

\subsection{TES thermal conductivity parameters}\label{thermal_params}
The phonon power, $\Pph$, is modeled as the following function of bath temperature $\Tb$, critical superconducting temperature $\Tc$, thermal conductivity constant $\kappa$, and thermal conductivity power law index $n$:

\begin{equation}
\Pph = \kappa(\Tc^n-\Tb^n).
\label{eqn:Pphi}
\end{equation}

Dark I--V curves acquired every $1\mK$ step between bath temperatures of $70\mK$ and $200\mK$ compose a dataset of $\Pph$ vs. $\Tb$ that is used to extract parameters $n$, $\Tc$, and $\kappa$. The focal plane temperature is recorded by a Rox thermometer and stabilized with a resistive heater whose current is continuously adjusted by a closed-loop servo system. We use indices $i$ or $j$ to denote an I--V curve acquired at a particular $\Tb$ . At each $\Tb$ we estimate the slope $\Delta\Pph/\Delta\Tb$ by computing $\Delta\Pph$ and $\Delta\Tb$ from the I--V curves acquired at $\Tb\pm d\Tb$ with $d\Tb = 1\mK$. The $\Pph$ vs. $\Tb$ slope is related to the thermal conductivity parameters through:

\begin{equation}
\left(\frac{\Delta \Pph}{\Delta \Tb}\right)_i = -\kappa n (\Tb)_i^{n-1}.
\label{eqn:dPphi}
\end{equation}

To isolate $n$ from $\kappa$ we compute the ratio of slopes at two different $\Tb$:

\begin{equation}
\frac{\left(\frac{\Delta \Pph}{\Delta \Tb}\right)_i}{\left(\frac{\Delta \Pph}{\Delta \Tb}\right)_j} = \frac{(\Tb)_i^{n-1}}{(\Tb)_j^{n-1}}.
\label{eqn:ratio_dPphi}
\end{equation}

\begin{figure} 
   \begin{center}
   \begin{tabular}{c} 
   \includegraphics[height=6.5cm]{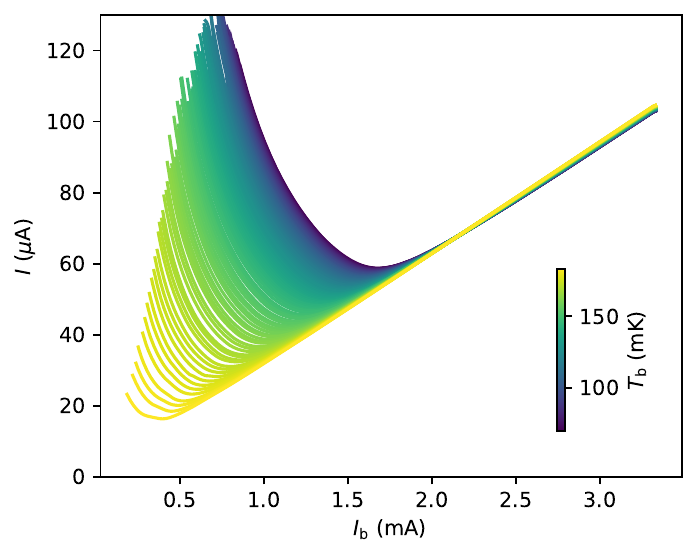}
   \includegraphics[height=6.5cm]{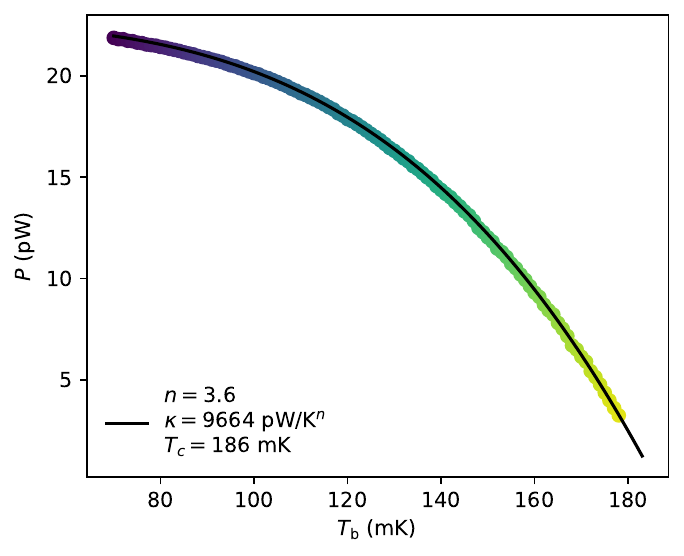}  
   \end{tabular}
   \end{center}
   \caption[I--V curves]
   { \label{fig:iv_G} 
\textbf{Left:}  plot of I--V data at multiple $\Tb$ for one detector. On the y-axis TES current $I$ and on the x-axis the bias current $\Ib$. \textbf{Right:}  plot of $\Psat$ vs. $\Tb$ results associated with the I--V curves on the left, along with the model fit to equation~\ref{eqn:Pphi} with parameters $n$, $\kappa$, and $\Tc$ .}
   \end{figure}

We start our analysis by constraining $n$, which captures the curvature of the $\Pph$ vs. $\Tb$ curve (see Figure~\ref{fig:iv_G}), and can be constrained for a single detector by the following logarithmic ratio of data points:

\begin{equation}
n = 1+\log \frac{\left(\frac{\Delta \Pph}{\Delta \Tb}\right)_i}{\left(\frac{\Delta \Pph}{\Delta \Tb}\right)_j}/ \log \frac{(\Tb)_i}{(\Tb)_j}.
\label{eqn:n}
\end{equation}

A pair of I--V curves acquired at $\Tb < \Tc$ provides a single estimate of $n$. With $\sim100$ I--V measurements below $\Tc$ in our dataset we obtain $\sim50$ independent measurements of $n$, which we average to obtain a single value for each detector. This method relies on high quality I--V data acquired across a wide range of $\Tb$ with small temperature spacing of $\sim\SI{1}{\milli\kelvin}$. Reliable sub-mK $\Tb$ thermometry and servo control is necessary to accurately measure $\Pph$ vs. $\Tb$.

For each detector we obtain an estimate of $n$. We average $n$ across all detectors in the array to obtain a single array averaged $n$. This single value is used for all further per-detector fitting to extract $\kappa$ and $\Tc$. Choosing a single array average $n$ allows us to easily compare TES thermal parameters across detectors. Any intrinsic variation of $n$ between detectors translates to additional variance in $\kappa$.
%See figure~?? showing the distribution of $n$ values for the CLASS-NIST detectors considered in this work.
$n$ is computed only from module 3 and module 4 data since module 1 and module 2 I--V data quality is affected by discrete jumps in $I$ that are under investigation. 
We find an average $n = 3.6$ with a mean error of $0.01$. This result is consistent with the \SI{100}{\milli\kelvin} SiN membrane ballistic phonon-transport measurements reported in \citenum{hoevers05}. 
With $n$ set to 3.6, $\kappa$ can be derived from the derivative of the $\Pph$ with respect to $\Tb$:

\begin{equation}
\kappa_i  = -\frac{1}{n}\left(\frac{1}{\Tb^{n-1}}\frac{\Delta \Pph}{\Delta \Tb}\right)_i .
\label{eqn:kappa}
\end{equation}
A per-detector $\kappa$ value is determined by averaging across all $\Tb$ measurements ($\kappa = \overline{\kappa_i}$). $\Tc$ is calculated per detector from:

\begin{equation}
(\Tc)_i  = \left(\frac{(\Pph)_i}{\kappa}+(\Tb)^n_i\right)^{1/n} .
\label{eqn:Tc}
\end{equation}
Again, the per-detector $\Tc$ is calculated from the average across the multiple $\Tb$ measurements ($\Tc$ = $\overline{(\Tc)_i}$). 

\begin{figure} 
   \begin{center}
   \begin{tabular}{c} 
   \includegraphics[height=6.28cm]{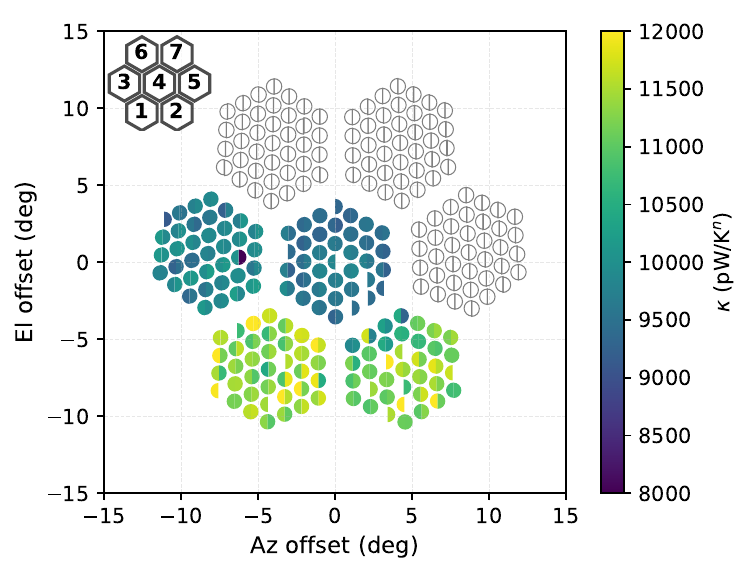}
   \includegraphics[height=6.28cm]{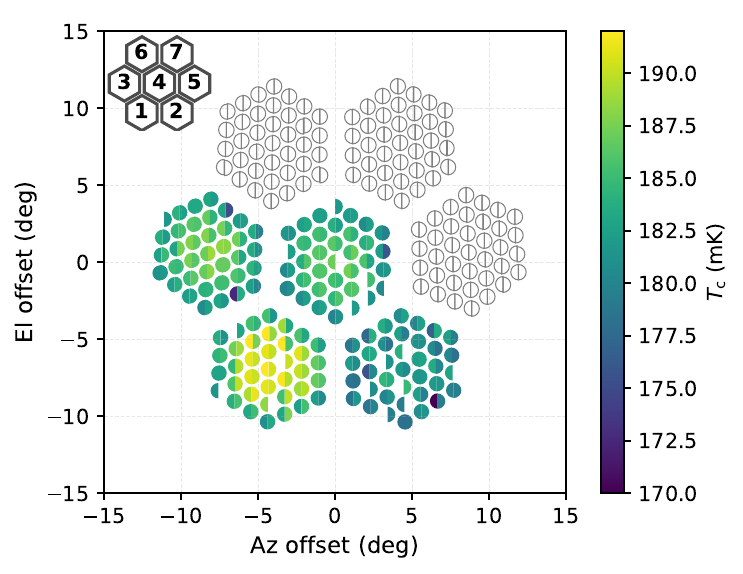}
   \\
   \includegraphics[height=6.28cm]{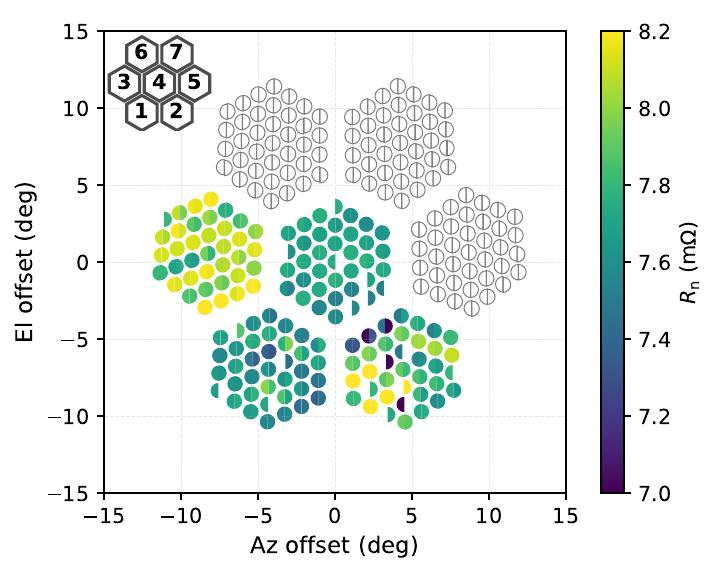} 
   \includegraphics[height=6.28cm]{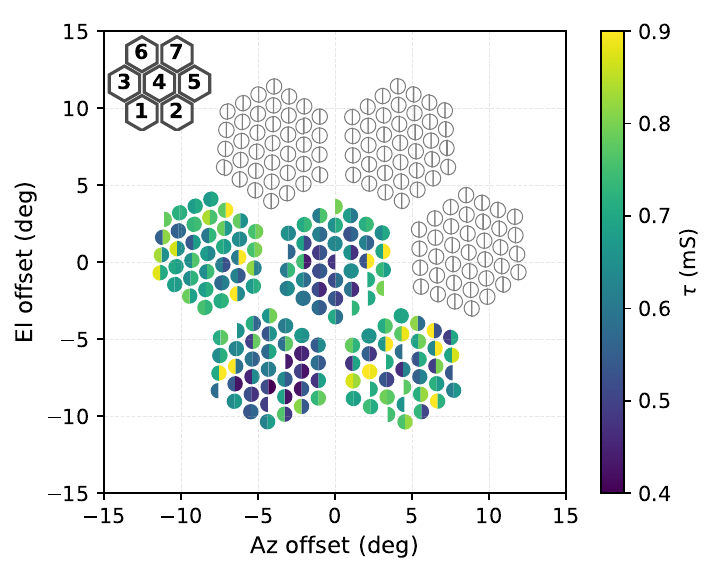}  
   \end{tabular}
   \end{center}
   \caption[Array plots of TES parameters] 
   { \label{fig:iv_param} 
\textbf{Top left:}  array plot of detector thermal conductivity constant $\kappa$. \textbf{Top right:}  array plot of detector critical temperature $\Tc$. \textbf{Bottom left:} array plot of TES normal Resistance $\Rn$. \textbf{Bottom right:}  array plot of on-sky detector time constant $\tau$ calculated from the hysteresis of the VPM synchronous signal. In each array plot, the paired half circles represent the two bolometers of a single feed. They are connected to the OMT antennas and sense polarizations oriented at $+45$ and $-45$ degrees. The color-filled half-circles correspond to the four deployed modules. The blank half-circles correspond to the location of three additional modules to be deployed in 2027.  }
   \end{figure}

See Table \ref{tab:tesparam_thermal} with average and 1$\sigma$ standard deviations of per-module parameters $\kappa$, $\Tc$, $n$, $G$, and $\Rn$. $\Rn$ is extracted from the TES resistance on the normal branch of the I--V curves. Figure \ref{fig:iv_param} shows that the per-detector values of $\Rn$ are generally uniform across the array, with the exception that module 3 exhibits slightly higher $\Rn$. The thermal conductivity $G$ of the TES is related to $\kappa$, $n$, and $\Tc$ through:

\begin{equation}
G = \left.\frac{d\Pph}{d \Tc}\right|_{\Tb} = \kappa n \Tc^{n-1}.
\label{eqn:G}
\end{equation}

Figure~\ref{fig:iv_param} shows array plots of the individual detector TES thermal parameters $\kappa$ and $\Tc$.  Modules 3 and 4 have a $\sim15\%$ lower average $\kappa$ than modules 1 and 2. We observe that both $\kappa$ and $\Tc$ are uniform across each module, although $\Tc$ shows a slight radial gradient, decreasing from the center toward the edges. 

%Discuss Tc target, offsets, etc
%Discuss kappa design, dimensions, similar to SO and cite. 
%Discuss G, higher than target, increases detectors noise, can reduce by ...
%Could add typical Ib, I, V, P, Psat values at operation to the table. 

\begin{table}
\caption[TES thermal and electrical parameters]{TES thermal and electrical parameters for optical detectors averaged per-module and using the module 3 and module 4 average $n=3.6$. $\Pph$ is computed assuming $\Tb = \SI{50}{\milli\kelvin} $. Along with the module average, we report the $1\sigma$ standard deviation of the distribution of individual detector values across the module. The second column gives the NIST fabrication number of the detector wafer in each module.    } 
\label{tab:tesparam_thermal}
\begin{center}       
\begin{tabular}{lllllll} 
\hline
\hline
\rule[-1ex]{0pt}{3.5ex}  Module & Wafer & $\Tc$ (\SI{}{\milli\kelvin}) & $\kappa$ (\SI{}{\pico\watt\per\kelvin^{n}}) & $G$ (\SI[per-mode = symbol]{}{\pico\watt\per\kelvin}) & $\Pph$ (\SI{}{\pico\watt}) & $\Rn$  (\SI{}{\milli\ohm}) \\
\hline
\rule[-1ex]{0pt}{3.5ex}  1 & 3 & $187\pm3$ & $11394\pm405$ & $524\pm26$ &  $27.0\pm1.7$ & $7.6\pm0.1$ \\
\rule[-1ex]{0pt}{3.5ex}  2 & 5 & $181\pm2$ & $11036\pm571$ & $465\pm27$ &  $23.2\pm1.5$ & $7.8\pm0.3$ \\
\rule[-1ex]{0pt}{3.5ex}  3 & 1 & $184\pm3$ & $9797\pm390$ & $433\pm24$ &  $22.0\pm1.3$ & $8.0\pm0.2$  \\
\rule[-1ex]{0pt}{3.5ex}  4 & 2 & $183\pm2$ & $9525\pm177$ & $417\pm18$ &  $21.1\pm1.2$ & $7.7\pm0.1$ \\
\rule[-1ex]{0pt}{3.5ex}  Array &  & $184\pm$3 & $10432\pm892$ & $460\pm47$ &  $23.3\pm2.7$ & $7.8\pm0.3$ \\
\rule[-1ex]{0pt}{3.5ex}  Target & & $160$ & $9841$ & $302$ &  $13.2$ & $8$ \\
\hline
\end{tabular}
\end{center}
\end{table}

\subsection{Responsivity}\label{responsivity}

The SQUID input, connected in series with the TES, measures changes in the TES current ($dI$) that are recorded by the MCE readout cards.
To combine data across multiple detectors and observing conditions, we determine the responsivity ($S$) calibration factors that convert a change $dI$ to a change in power dissipated at the TES ($dP$). For CLASS, responsivity is determined from I--V data acquired before the observed time-ordered data (TOD).  A per-detector optical efficiency factor (determined from planet observations) combined with bandpass measurements provides a final calibration from $dP$ to sky optical power $d\Popt$ or equivalently to CMB temperature $d\Tcmb$.

Three I--V responsivity estimates are considered for calibrating on-sky data, all proportional to the TES shunt resistor $\Rshunt$. For CLASS we use shunt chips fabricated by NIST with $\Rshunt = \SI{250}{\micro\ohm}$. 
%We assume the design shunt resistance value for the analysis ahead. 
We have no indication the shunt resistance changes across time and qualifying tests of the shunt chips show their resistances match the design value closely. Any small variation of shunt resistances across the array are expected to be constant across time, and therefore are absorbed into the per-detector optical efficiency calibration. 

The simplest $\Ib$ responsivity estimate is given by:

\begin{equation}
S_{\Ib}^{-1} =\left.\frac{dP}{dI}\right|_{I_{\mathrm{b}}} =  -\Rshunt \Ib.
\label{eqn:respIb}
\end{equation}
$\Ib$ is always known since we set it to a value that places the detectors on the superconducting transition. Computing $\Ib$  responsivity does not require I--V data, but note that in practice, we use I--V data to choose the optimal $\Ib$ target.  The systematic error of $\Ib$  responsivity is typically in the tens of percent range, resulting in an over-estimate of the detector noise and an under-estimate of the optical efficiency. A second improved responsivity estimate ($S^*$) with reduced systematic error incorporates knowledge of the absolute  TES current $I$ at the operating bias:
\begin{equation}
(S^*)^{-1} =\left.\frac{dP}{dI}\right|_{I_{\mathrm{b}}} =  -\Rshunt (\Ib-2I).
\label{eqn:resp_star}
\end{equation}
The absolute TES current $I$ at the operating bias can be extracted per-detector from I--V data acquired at the same $\Tb$ temperature and under similar optical loading. The $S^*$ responsivity is derived from the ideal condition that the TES $P$ is constant throughout the transition.

A third more accurate responsivity estimate described in Ref.~\citenum{appel22} relies on the slope $\Delta I_{\mathrm{b}}/{\Delta I}$ of the I--V data:
\begin{equation}
    S^{-1} = R_{\mathrm{sh}}I\frac{\Delta I_{\mathrm{b}}}{\Delta I}
    \left(1-\mathscr{L}_I^{-1}\right),
    \label{eqn:resp_slope}
\end{equation}
where the loop gain $\mathscr{L}_I$ is defined as:
\begin{equation}
    \Lg = \frac{P \alpha}{G T} =  \frac{P}{G T} \left.\frac{T}{R}\frac{\partial R}{\partial T}\right|_{I}
    = \frac{I^2}{G}\left.\frac{\partial R}{\partial T}\right|_{I}.
    \label{eqn:loop_gain}
\end{equation}
Typically $\mathscr{L}_I$ is large at the operating bias, and we approximate Equation~\ref{eqn:resp_slope} with:
\begin{equation}
    S^{-1} \approx R_{\mathrm{sh}}I\frac{\Delta I_{\mathrm{b}}}{\Delta I}.
    \label{eqn:resp_slope_approx}
\end{equation}
To reduce errors in the I--V responsivity estimate we implement the I--V bin method described in Ref.~\citenum{appel22}. Responsivity results from I--V curves acquired under similar observing conditions across all observing seasons are binned and averaged, yielding an improved responsivity for each scan set. 

Table~\ref{tab:tesparam_s} summarizes the module average responsivity and optical time constants from on-sky observations. Three responsivity estimates are included $S$, $S^*$, and $S_{\Ib}$.  For module 3 and module 4, $S \approx S^*$ as expected from a TES model with high $\Lg$ when biased lower on the transition.  For module 1 and module 2,  $S^{-1} > (S^*)^{-1}$ by $\sim20\%$. This difference is also observed in the dark multi-$\Tb$ I--V $S$ described later in this section. The cause of this unexpected difference in responsivities is under investigation, although it appears to be related to the bias power $P$ of the TES increasing as one decreases the operating TES resistance. For the NEP and NET analysis in the following sections we use the I--V slope responsivity $S$ for all four modules. The optical loading analysis is derived from I--V data directly; therefore it does not depend on the responsivity calibration. 

During dark lab detector tests we acquire a limited set of I--V data not suitable for the I--V bin method. On the other hand, the dark multi-$\Tb$ temperature datasets provide a fourth method to compute responsivity. This accurate responsivity estimate is determined from the change to TES current $(\Delta I)_{ij}$ and bias power $(\Delta P)_{ij}$ corresponding to a change in $(\Delta \Tb)_{ij}$ between I--V curves $i$, and $j$:

\begin{equation}
    S^{-1} = \frac{\Delta P_{ij}}{\Delta I_{ij}}.
    \label{eqn:resp_Tb}
\end{equation}
The I--V pairs $i$, $j$ are chosen such that $(\Delta \Tb)_{ij}$ is big enough to make $\Delta P_{ij}$ larger than the measurement noise, but at the same time small enough that the amplitude of $\Delta I_{ij}$ is close to the small signal limit of our CMB observations. Multiple pairs of I--V curves in the dark $\Tb$ datasets satisfy the previous conditions, and are averaged to obtain a single responsivity estimate. 

\begin{figure} 
   \begin{center}
   \begin{tabular}{c} 
   \includegraphics[height=6.5cm]{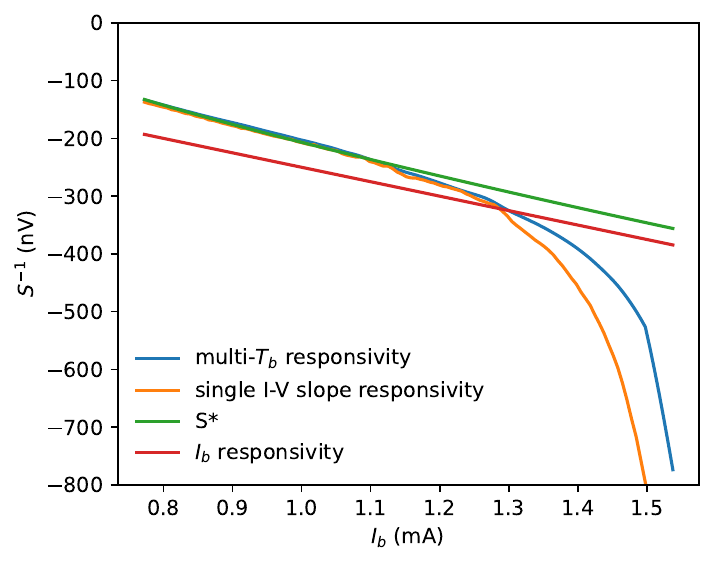}
   \includegraphics[height=6.5cm]{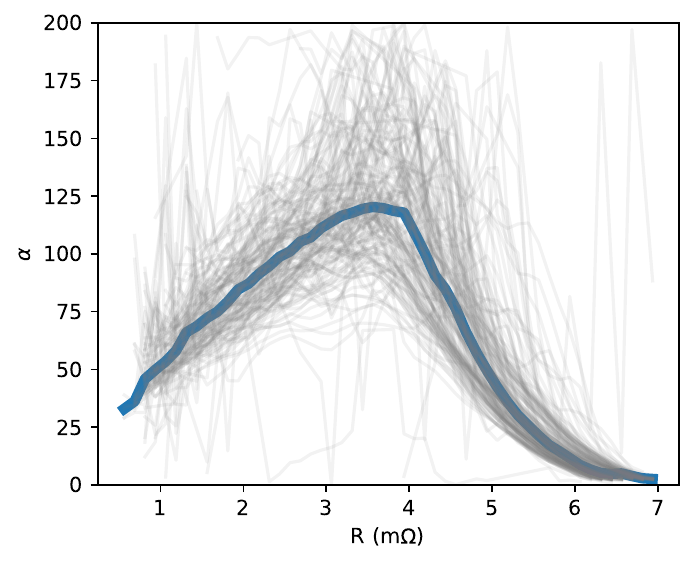}
   \end{tabular}
   \end{center}
   \caption[TES responsivity and $\alpha$] 
   { \label{fig:resp_alpha} 
\textbf{Left:}  single detector example of the four different I--V curve derived responsivity estimates at $\Tb = \SI{110}{\milli\kelvin}$ and $\Popt=0$. The robust $\Ib$ responsivity is systematically offset low from the other estimates when the TES bias is set below $\SI{1.3}{\milli\ampere}$. As expected at lower $\Ib$ targets, the multi-$\Tb$ and I--V slope responsivities match $S^{*}$, which corresponds to the ideal case when $\Pph$ is constant along the transition\cite{appel22}. The I--V slope responsivity (Equation~\ref{eqn:resp_slope_approx}) deviates from the more accurate multi-$\Tb$ responsivity (Equation~\ref{eqn:resp_Tb}) at higher $\Ib$, where the $\Lg \gg 1$ approximation is no longer valid. \textbf{Right:}  plot of module 3 and module 4 averaged $\alpha$ estimated from the dark lab I--V dataset. Over-plotted in gray are the individual detector $\alpha$ estimates.}
   \end{figure}

Figure~\ref{fig:resp_alpha} shows the multiple responsivity estimates for a single module 3 TES detector. Note that at medium and low bias current setpoints the multi-$\Tb$, I--V slope responsivity $S$, and $S^{*}$ match. When biased higher on the transition $S^{-1}$ and $\Lg$ drop; in this range multi-$\Tb$ responsivity provides the most accurate method, while I--V slope responsivity is biased due to the approximation of Equation~\ref{eqn:resp_slope_approx}.

\begin{table}
\caption[TES responsivity]{On-sky small signal limit TES responsivity ($S$) and time constant ($\tau$) averaged per-module and across an observing season.  Three responsivity estimates are included: I--V slope responsivity equation~\ref{eqn:resp_slope_approx}, $S^{*}$ responsivity equation~\ref{eqn:resp_star}, and $\Ib$ responsivity equation~\ref{eqn:respIb}.  The row labeled `Array' is the average of all four modules.  } 
\label{tab:tesparam_s}
\begin{center}       
\begin{tabular}{lllll} 
\hline
\hline
\rule[-1ex]{0pt}{3.5ex}  Module & $S^{-1}$ (\SI{}{\nano\volt}) & $(S^*)^{-1}$ (\SI{}{\nano\volt}) & $S_{\Ib}^{-1}$ (\SI{}{\nano\volt})  & $\tau_{\gamma}$ (\SI{}{\milli\second})  \\
\hline
\rule[-1ex]{0pt}{3.5ex}  1 & $-182\pm14$ & $-225\pm14$ & $-271\pm13$  & $0.61\pm0.12$ \\
\rule[-1ex]{0pt}{3.5ex}  2 & $-188\pm18$ & $-220\pm15$ & $-261\pm15$  & $0.71\pm0.16$ \\
\rule[-1ex]{0pt}{3.5ex}  3 & $-215\pm19$ & $-212\pm17$ & $-251\pm15$  & $0.72\pm0.14$  \\
\rule[-1ex]{0pt}{3.5ex}  4 & $-189\pm19$ & $-185\pm18$ & $-228\pm17$  & $0.63\pm0.12$ \\
\rule[-1ex]{0pt}{3.5ex}  Array & $-194\pm22$ & $-211\pm22$ & $-253\pm22$  & $0.67\pm0.14$ \\
\hline 
\end{tabular}
\end{center}
\end{table}

\subsection{TES loop gain}\label{loop_gain}

The responsivity calculated from the dark multi-$\Tb$ dataset (Equation~\ref{eqn:resp_Tb}) can be divided by I--V slope responsivity from Equation~\ref{eqn:resp_slope_approx} to obtain a measurement of the TES loop gain $\Lg$:

\begin{equation}
\left(1-\Lg^{-1}\right) =  \frac{\Delta P_{ij}}{\Delta I_{ij}}\frac{1}{R_{\mathrm{sh}}I}\frac{\Delta I}{\Delta I_{\mathrm{b}}}.
\label{eqn:loop_gain_meas}
\end{equation}

Equation~\ref{eqn:loop_gain} may be used to convert $\Lg$ to the TES parameter $\alpha$. The average $\alpha$ of module 3 and module 4 extracted from the multi-$\Tb$ I--V dataset is shown in Figure~\ref{fig:resp_alpha} plotted as a function of TES resistance. During observations, the TESs are biased near the middle of the transition where $\alpha\approx100$. This estimate of $\alpha$ relies on accurate I--V curve measurements at multiple bath temperatures. The quality of the module 3 and module 4 multi-$\Tb$ datasets yields good results. The module 1 and module 2 I--V data suffer from small TES current jumps, which add excess variance to the I--V slope measurement, not allowing an accurate estimate of $\Lg$.  This excess module 1 and module 2 I--V curve noise is under investigation, but it does not appear to affect our standard observing TOD data acquired at a single $\Ib$. Inputting the array average detector parameters ($P = \Pph-\Popt \approx\SI{20}{\pico\watt}$, $\alpha\approx100$, $G = \SI[per-mode = symbol]{460}{\pico\watt\per\kelvin}$, and $\Tc = \SI{0.184}{\kelvin}$ ) into Equation~\ref{eqn:loop_gain} yields an array average $\Lg \approx 24$.

\subsection{Time constant}\label{time_constant}

The response of the TES bolometer to a time varying optical signal is modeled with a single-pole transfer function with an optical time constant $\tau_{\gamma}$. The CLASS polarization modulators place the sky polarization near 10~Hz; therefore to maximize the signal-to-noise ratio we desire detectors with $\tau_{\gamma} \ll 1/(2\pi\times\SI{10}{\hertz}) = \SI{16}{\milli\second}$.
By fitting the hysteresis of the VPM synchronous signal\cite{appel19} across the CLASS-W2 2025 dataset, we extract per-detector median time constants. The average detector time constants across modules are summarized in Table~\ref{tab:tesparam_s}. The median array time constant of \SI{0.67}{\milli\second} achieves the desired detector speed goal. Figure \ref{fig:iv_param} shows that the per-detector time constants are similar across the array.

The optical response of the TES bolometer is determined by the electro-thermal speed up of the bolometer's thermal time constant $\tau = C/G$, where $C$ is the heat capacity of the TES bolometer island. The CLASS design target of $C =\SI{3}{\pico\joule\per\kelvin}$, together with the array median thermal conductivity $G = \SI{460}{\pico\watt\per\kelvin}$ imply a thermal time constant of $\tau = \SI{6.5}{\milli\second}$. Therefore, the TES electro-thermal feedback speeds up the thermal time constant $\tau$ by a factor of $\sim$10. 
Using the TES time constant model in \citenum{appel22} the array average speedup factor, equation~\ref{eqn:tau_opt}, is consistent with $\Lg \approx 24$, and $\beta\approx1.5$, where $\beta$ is the logarithmic current sensitivity of the superconducting transition $\beta = \frac{I}{R}\frac{dR}{dI}$.

\begin{equation}
\frac{\tau_{\gamma}}{\tau} = \frac{1+\beta+R_{\mathrm{sh}}/R}
{1+\beta+R_{\mathrm{sh}}/R+\left(1-R_{\mathrm{sh}}/R\right)\mathscr{L}_I}.
\label{eqn:tau_opt}
\end{equation}

\section{Detector Optical characterization}\label{sec:tes_param_opt}

The sensitivity, noise, and optical response of the CLASS-W2 detectors were determined using on-sky observations, while the detector bandpass was measured in the lab with a Fourier transform spectrometer (FTS). This section estimates the detector array mapping speed and summarizes the measured bandpass and beam response, both critical inputs for interpreting the final sky maps.

\begin{table}
\caption[Detector bandpass]{TES bandpass parameters per-module, including effective center frequencies for beam filling sources with the following spectra: Rayleigh–Jeans ($\frj$), CMB ($\fcmb$) blackbody, Galactic synchrotron ($\fsync$), and Galactic dust $\fdust$. The bandwidth $\Delta \nu$ is computed from the difference in frequency between the two half-power points of the bandpass edges. The conversion factor $\frac{d\Trj}{d\Popt}$ from power to  Rayleigh–Jeans temperature assumes detector optical efficiency of one. The derivative $\frac{d\Tcmb}{d\Trj}$ gives the factor that converts Rayleigh–Jeans temperature fluctuations into equivalent CMB temperature fluctuations. The fabricated bandpass $\fc$ is shifted high by $\sim\SI{3}{\giga\hertz}$ and has a smaller bandwidth when compared to the simulated design. } 
\label{tab:tesparam_bp}
\begin{center}       
\begin{tabular}{cccccccc} 
\hline
\hline
\rule[-1ex]{0pt}{3.5ex}  Module & $\frj$ (\SI{}{\giga\hertz})  & $\fcmb$ (\SI{}{\giga\hertz}) & $\fsync$ (\SI{}{\giga\hertz}) & $\fdust$ (\SI{}{\giga\hertz}) & $\Delta \nu$ (\SI{}{\giga\hertz})  & $\frac{d\Trj}{dP_{\gamma}} (\SI{}{\kelvin\per\pico\watt})$ & $\frac{d\Tcmb}{d\Trj}$ \\
\hline
\rule[-1ex]{0pt}{3.5ex}  1 & 94.7 & 93.8 & 92.1 & 95.9 & 28.9 & 2.51 & 1.25 \\
\rule[-1ex]{0pt}{3.5ex}  2 & 95.3 & 94.5 & 92.9 & 96.5 & 28.0 & 2.59 & 1.25 \\
\rule[-1ex]{0pt}{3.5ex}  3 & 95.4 & 94.5 & 93.0 & 96.6 & 28.1 & 2.58 & 1.25 \\
\rule[-1ex]{0pt}{3.5ex}  4 & 95.5 & 94.7 & 93.1 & 96.7 & 28.1 & 2.58 & 1.25 \\
\rule[-1ex]{0pt}{3.5ex}  Array & 95.2 & 94.4 & 92.8 & 96.4 & 28.3 & 2.56 & 1.25 \\
\rule[-1ex]{0pt}{3.5ex}  Sim & 92.6 & 91.7 & 89.7 & 94.1 & 31.0 & 2.34 & 1.24 \\
\hline 
\end{tabular}
\end{center}
\end{table}

\subsection{Bandpass}\label{bandpass}
   \begin{figure} 
   \begin{center}
   \begin{tabular}{c} 
   \includegraphics[height=6.5cm]{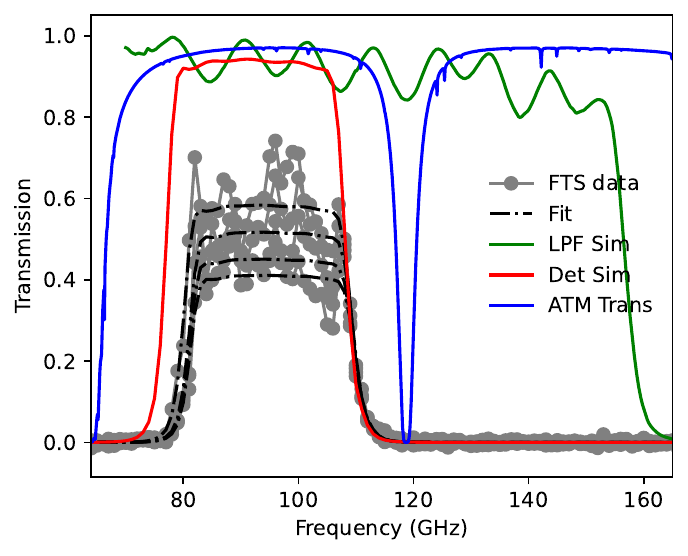}
   \includegraphics[height=6.5cm]{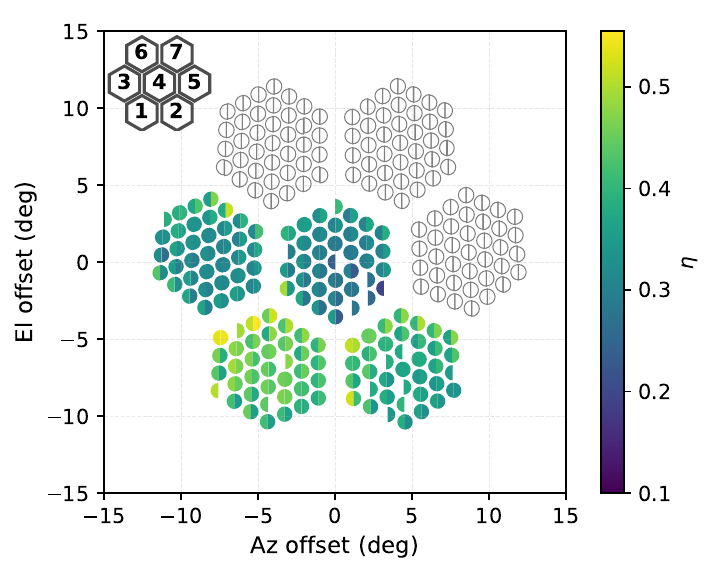}
	\end{tabular}
	\end{center}
   \caption[Detector bandpass and optical efficiency] 
   { \label{fig:bandpass} 
\textbf{Left:}  FTS bandpass measurements per-module plotted in gray~{\color{gray}\rule[0.5ex]{0.3cm}{2pt}}, while in black~{\color{black}\rule[0.5ex]{0.3cm}{2pt}} we plot fits to the measurements that average the in-band bandpass fluctuations. The transmission amplitudes of the per-module bandpass measurements are normalized to the detector-only optical efficiency, derived from the on-sky end-to-end optical efficiency reported in Table~\ref{tab:tesparam_opt} divided by the estimated telescope efficiency of $\eta_t = 0.71$ . The lower-band edge is at \SI{81.3}{\giga\hertz} except for module 1 at \SI{80.3}{\giga\hertz}. The upper-band edge is at \SI{109.4}{\giga\hertz}. The targeted simulated bandpass shown in red~{\color{red}\rule[0.5ex]{0.3cm}{2pt}} has a center frequency about \SI{3}{\giga\hertz} below the measured bandpass. The 50th percentile atmospheric transmission at the CLASS site\cite{paine24} is shown in blue~{\color{blue}\rule[0.5ex]{0.3cm}{2pt}}. Plotted in green~{\color{teal}\rule[0.5ex]{0.3cm}{2pt}} is the simulated LPF bandpass with a half-maximum cutoff frequency of \SI{157}{\giga\hertz}. \textbf{Right:}  CLASS-W2 array plot of the end-to-end detector optical efficiency from 2025 Jupiter observations. The location of each circle corresponds to the feedhorn elevation and azimuth offset angle pointing with respect to boresight. The two half circles correspond to the two bolometers in each feed connected to the OMT antennas aligned at $+45$ and $-45$ degrees. Module 1 and module 2, achieve higher optical efficiency than module 3 and module 4. The unfilled circles correspond to the location of the three additional modules scheduled to be deployed in 2027.  }
   \end{figure} 

The CLASS-W2 detector frequency bandpass $f(\nu)$ set by the on-chip band-defining filters was measured in-lab with a Martin--Puplett FTS\cite{mp_fts} illuminating the detector focal plane. In addition to the standard CLASS telescope cryogenic lenses and IR filters, we installed a neoprene neutral density filter (NDF) at \SI{4}{\kelvin} and a polarizer at the window to reduce the \SI{300}{\kelvin} optical load from the lab environment. The polarizer spectral response is flat in the detector band, while the red-transmission spectrum of the neoprene NDF is characterized by the ratio of the bandpass measurement with and without a second neoprene NDF mounted outside the window.

Compared to the initial simulated on-chip filter bandpass, the FTS measured bandpass edges are shifted high by \SI{4}{\giga\hertz} on the lower end and \SI{1}{\giga\hertz} on the higher end. These shifts reduce the achieved bandwidth by $\sim3$~GHz compared to the design target, but are small enough that the detector bandpass still falls comfortably within the \SI{90}{\giga\hertz} atmospheric observing window.
 The cause of the observed bandpass shift is still under investigation, but could be the result of a shift in the relative permittivity of the SiN dielectric and/or a significant over-etch of the microstrip \cite{jaehnig26}. 
 Accounting for the observed bandwidth reduction would require a microstrip over-etch far greater than even the worst-case fabrication tolerance.
Future detector fabrication runs will enhance detector sensitivity by recovering a $\sim$10\% increase in bandwidth. This can be accomplished by updating the on-chip filter design and through improved control of the dielectric properties.

If a source fully fills the beam and each detector couples to a single mode, the optical throughput is $A \Omega = \lambda^{2}/2$, where $\Omega$ is the detector beam solid angle and $A$ is the collecting area. Table~\ref{tab:tesparam_bp} summarizes the  brightness temperature center frequencies per module for the following types of sources: 1. Rayleigh–Jeans ($\frj$) flat spectrum $\sigma(\nu) = 1$; 2. CMB ($\fcmb$) blackbody $\Tcmb=\SI{2.725}{\kelvin}$,   $\sigma(\nu) = x/(\exp{x}-1)$ where $x = h \nu/(\kb \Tcmb)$, $h$ is the Planck constant and $\kb$ is the Boltzmann constant; 3. Galactic synchrotron ($\fsync$) emission power law $\sigma(\nu) =\nu^{-3.1}$ \cite{planck15IV}; and 4. Galactic dust ($\fdust$) emission power law  $\sigma(\nu) =\nu^{1.55}$ \cite{planck15IV}.
The center frequency of the bandpass ($\fc$) for a source with frequency dependence  $\sigma(\nu)$ is given by:

\begin{equation}
    \fc = \frac{\int\nu f(\nu)\sigma(\nu)\, d\nu }{\int f(\nu) \sigma(\nu)\, d\nu}.
    \label{eqn:fc}
\end{equation}

The optical power $\Popt$ coupled onto one of our single-mode bolometers (assuming perfect optical efficiency) from a beam filling  Rayleigh–Jeans source at temperature $\Trj$ is $\Popt= \kb \Delta \nu \Trj$. The conversion factor from sky power to temperature $\frac{d\Trj}{d\Popt} = (\kb \Delta \nu)^{-1}$ is given for each module average bandpass in Table~\ref{tab:tesparam_bp}, as well as the temperature conversion factor from a Rayleigh–Jeans source to the CMB blackbody $\frac{d\Tcmb}{d\Trj}= (\exp{x}-1)^{2}/(x^{2}\exp{x})$ where $x = h \nu/(\kb \Tcmb)$.
The bandwidth ($\Delta\nu$) is calculated from the difference between the low-frequency half-power point to the high-frequency half-power point of the bandpass. This is a robust estimate of the bandwidth, which is less susceptible to FTS measurement systematic errors that show up as fluctuations in the in-band bandpass response.
Averaging all detector FTS bandpass measurements constrains average out-of-band optical coupling at the $<0.2\%$ level. 
%The FTS probes direct out-of-band coupling through the OMT antennas. Alternatively, broadband power coupled directly to the TES bolometer island may not be fully constrained by our FTS measurements.

\subsection{Optical efficiency}\label{optical_efficiency}

%The end-to-end detector optical efficiency ($\eta$), including the telescope optics, is obtained by computing the ratio of the measured point source signal power ($P$) to the expected power ($\Ps$): $\eta = P/\Ps$.  The observed amplitude is extracted from the TOD by fitting a model based on the detector beam profile, and calibrating the data to pico-watt power units with I--V responsivity. The expected source power $\Ps$ depends on the solid angle of the source ($\Omega_s$) and its brightness temperature $\Ts$. The brightness temperatures of Jupiter, Venus, and the Moon,the sources most commonly observed with CLASS,lie in the Rayleigh–Jeans regime; therefore:

%\begin{equation}
%\Ps \approx \frac{\Omega_s}{\Omega}\kb \Delta \nu \Ts 
%\label{eqn:Ps}
%\end{equation}

The end-to-end detector optical efficiency ($\eta$), including the telescope optics, is obtained by comparing the measured point source signal power ($P_{\mathrm{s}}$) to that predicted for an ideal optical system. The observed amplitude is extracted from the TOD by fitting a model based on the detector beam profile, and calibrating the data to pico-watt power units with I--V responsivity. The received power depends on the optical efficiency, the solid angle of the source ($\Omega_s$), and its brightness temperature $\Ts$. The brightness temperatures of Jupiter and Venus ---the point sources most commonly observed with CLASS---lie in the Rayleigh–Jeans regime; therefore:

\begin{equation}
P_{\mathrm{s}} = \eta \frac{\Omega_s}{\Omega}\kb \Delta \nu \Ts.
\label{eqn:Ps}
\end{equation}

The per-detector telescope optical efficiency was determined from 2025 Jupiter observations before the low-pass filter was installed. This analysis follows the same calibration procedure described in Ref.~\citenum{datta24}.
The mean array telescope optical efficiency was 0.37, while module 1 had a higher average optical efficiency of 0.44 with six detectors above 0.5 (see Figure~\ref{fig:bandpass}).  We isolate the detector optical efficiency ($\eta_d$) by dividing out the telescope optical efficiency $\eta_t = 0.71$ \cite{nunez25} that includes beam spill, as well as reflection and absorption of the multiple telescope optical elements such as filters, lenses, and window. The array average detector-only optical efficiency is $0.52$, and the average for module 1 is $0.61$.  The six detectors on the higher end of the optical efficiency distribution achieve a detector-only optical efficiency of $0.7$. The lower optical efficiency of detectors in module 3 and module 4 is under investigation. Nevertheless, all modules achieve photon noise limited performance. 

The LPF transmission model predicts $\nlpf = 0.93$ within the CLASS \SI{90}{\giga\hertz} bandpass. Moon observations constrain  $\nlpf \approx 0.98$, but with systematic uncertainty associated with: 1.~detector non-linear response to the large Moon signal, and 2.~uncertainty in our \SI{90}{\giga\hertz} Moon brightness temperature versus Moon phase model.
An updated optical efficiency and beam model will be constructed in 2027 once Jupiter enters our observing sky area again. The Jupiter antenna temperature is small, well within the small-signal limit of our TES bolometers. Moreover, Jupiter brightness temperature is well characterized at \SI{90}{\giga\hertz} by measurement from WMAP~\cite{weiland11} and Planck~\cite{planckLII} satellites. 
   \begin{figure} 
   \begin{center}
   \begin{tabular}{c} 
   \includegraphics[height=5.9cm, trim={0.4cm 0 0.4cm 0},clip ]{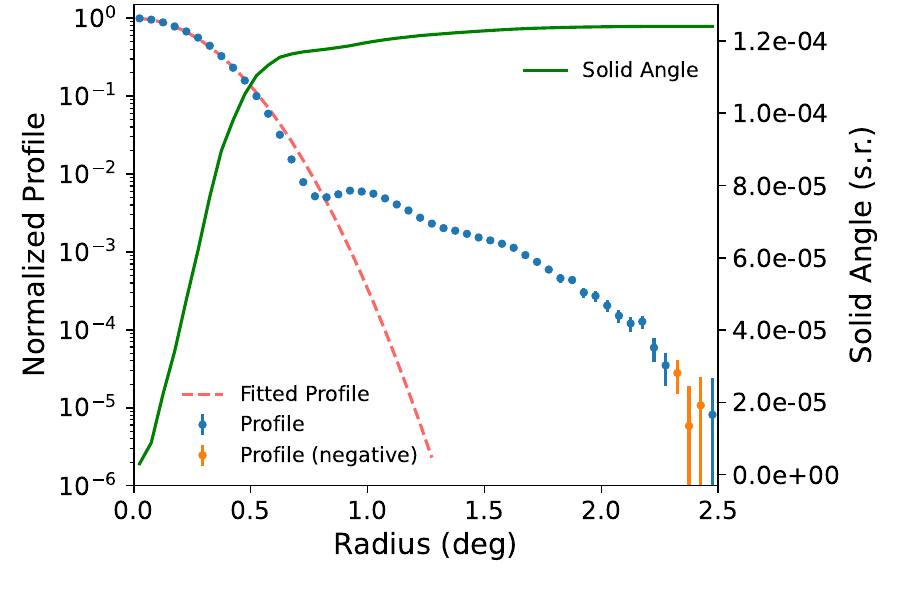}
   \includegraphics[height=5.9cm, trim={1.0cm 0 0.6cm 0},clip ]{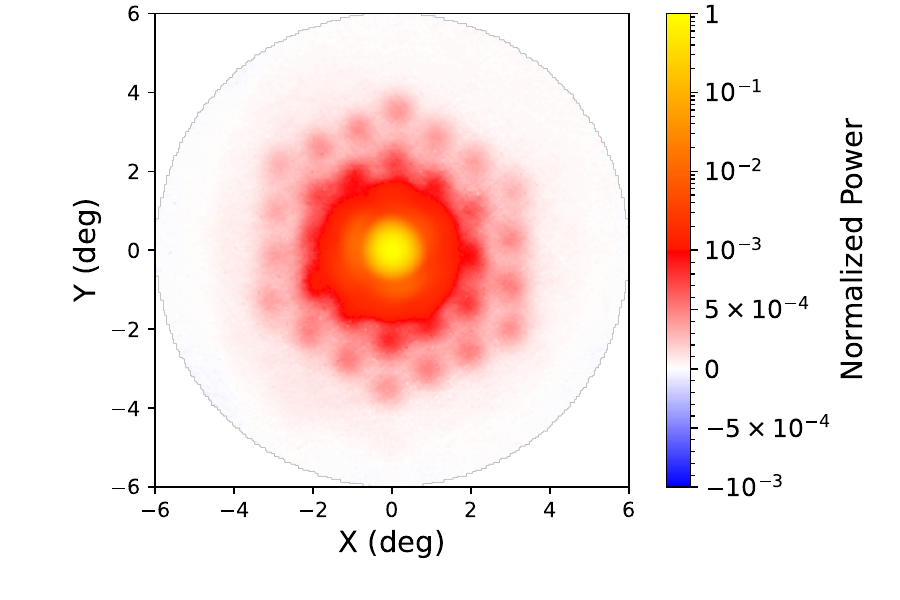}
	\end{tabular}
	\end{center}
   \caption[Beam maps] 
   { \label{fig:det_beam} 
\textbf{Left:}  2025 azimuthally averaged CLASS-W2 beam profile from measuring Jupiter in blue (FWHM = $\SI{0.592}{\deg}$), the beam cumulative solid angle in green ($\SI{124}{\micro\steradian}$) and a Gaussian beam profile fit in dashed red. The fit to a Gaussian beam is performed out to \SI{0.6}{\degree}. \textbf{Right:}  Stacked map of the Moon with LPF installed showing $<$0.1\% optical cross-talk signals from neighboring feedhorns in the same module. The amplitudes of the optical cross-talk signals decreased by 60\% after the LPF was installed.}
   \end{figure}

\subsection{Beam maps}\label{beam_maps}
Beam maps constructed from Jupiter observations in 2025 indicate an array average beam solid angle of $\Omega = \SI{124\pm0.3}{\micro\steradian}$, and a Full-Width-Half-Maximum (FWHM) angle of $\SI{0.592\pm0.003}{\deg}$ matching the expected beam profile \cite{datta24}. The detector beam profile measurement is shown in Figure~\ref{fig:det_beam}.

High signal-to-noise observations of the Moon show optical cross-talk signals replicating the module feedhorn offsets at the $<0.1\%$ signal level.
With the LPF in place, the amplitude of the optical cross-talk signal in the 2026 Moon observations drops by roughly $\sim$60\% relative to the 2025 data taken without the LPF filter, implying that a significant share of the optical cross-talk originates from the out-of-band `blue-leak´ power identified in the optical-loading analysis that follows. A 2026 Moon map for the feedhorn located at the center of a module is shown in Figure~\ref{fig:det_beam}. To produce this map, we averaged the data from the detectors in the central feedhorns of modules 3 and 4.

\subsection{Optical loading}\label{sec:optical_loading}
\begin{table}
\caption[Detector optical efficiency and optical loading]{Detector optical efficiency and optical loading per-module. The optical efficiencies, including detector and telescope contributions, were extracted from 2025 Jupiter observations. The dark-detector optical loading $\Pdd$ decreased with the LPF installed. This is clear evidence for the blue-leak model of high-frequency optical power coupling directly to the TES. LPF 2026 data were acquired in the austral summer with a higher average PWV ($\SI{1.5}{\milli\meter}$) than the 2025 no-LPF dataset ($\SI{0.8}{\milli\meter}$), mostly acquired in the austral winter. This partly explains the increase in $\Pinband$ between no-LPF and LPF. } 
\label{tab:tesparam_opt}
\begin{center}       
\begin{tabular}{ll|lll|lll} 
\hline
\hline
\rule[-1ex]{0pt}{3.5ex}   &  &  \multicolumn{3}{|c|}{no-LPF} & \multicolumn{3}{c}{LPF} \\
\hline
\rule[-1ex]{0pt}{3.5ex}  Module & $\eta$  & ${\Pod}$ (\SI{}{\pico\watt}) & $\Pdd$ (\SI{}{\pico\watt}) & $\Pinband$ (\SI{}{\pico\watt}) & $\Pod$ (\SI{}{\pico\watt}) & $\Pdd$ (\SI{}{\pico\watt}) & $\Pinband$ (\SI{}{\pico\watt})\\
\hline
\rule[-1ex]{0pt}{3.5ex}  1 & $0.44\pm0.05$ & $4.8\pm1.3$ & $1.1\pm0.8$ & $3.5\pm1.0$  & $4.3\pm1.1$ & $0.1\pm0.5$ & $4.1\pm1.1$\\
\rule[-1ex]{0pt}{3.5ex}  2 & $0.39\pm0.05$ & $4.5\pm1.5$ & $1.6\pm0.9$ & $3.0\pm1.1$  & $3.9\pm1.4$ & $0.3\pm0.6$ & $3.6\pm1.3$\\
\rule[-1ex]{0pt}{3.5ex}  3 & $0.34\pm0.04$ & $3.6\pm0.8$ & $1.1\pm0.6$ & $2.4\pm0.5$  & $3.1\pm0.8$ & $0.2\pm0.1$ & $2.9\pm0.8$ \\
\rule[-1ex]{0pt}{3.5ex}  4 & $0.31\pm0.05$ & $3.2\pm0.9$ & $0.8\pm0.8$ & $1.9\pm0.6$  & $2.7\pm0.8$ & $0.2\pm0.2$ & $2.4\pm0.8$\\
\rule[-1ex]{0pt}{3.5ex}  Array & $0.37\pm0.07$ & $4.1\pm1.3$ & $1.1\pm0.8$ & $2.7\pm1.0$  & $3.5\pm1.2$ & $0.2\pm0.4$ & $3.3\pm1.2$ \\
\hline 
\end{tabular}
\end{center}
\end{table}

The TES thermal conductivity parameters discussed in section \ref{thermal_params} determine $P_{\phi}$ (see Equation~\ref{eqn:Pphi}), equivalent to $\Psat$ with zero optical loading ($\Popt = 0$).  When observing the sky $\Psat$ drops by $\Popt$: 

\begin{equation}
\Psat = P_{\phi}-\Popt.
\label{eqn:Pgamma}
\end{equation}

In the ideal case $\Popt$ is entirely composed of radiation in the detector bandpass frequency range that is coupled through the OMT antennas.  We find that a small fraction of the optical power appears to couple directly to the TES bolometer island across a wide frequency range illustrated by the orange curve in Figure~\ref{fig:Pdd} (left); often this excess power is referred to as a `blue-leak', since a majority comes at frequencies above the detector bandpass. The waveguide connecting the feedhorn to the OMT has a cutoff frequency that efficiently attenuates optical power below the instrument’s bandpass.

This blue-leak power is measured by tracking the changes in saturation power of the dark TES detectors (DD). Ten DD are distributed across a module, and they share the same basic design as the optical detector (OD) TES but are not connected to any microstrips or antennas. Optical power coupling directly to the TES bolometer island is expected to be similar between OD and DD. Light that travels through the feedhorns can leak through the small gap between the detector wafer and the backshort (see Figure~\ref{fig:det_stack}), and a fraction of this optical power will dissipate directly on the TES bolometer island.  
We assume a simple optical loading model for the OD with a component of power coupled through the OMT antenna $\Pinband$, and a blue-leak component that couples directly to the TES $\Pbl$ :

\begin{equation}
\Pinband = \Popt-\Pbl.
\label{eqn:Pod}
\end{equation}

The total optical loading $\Popt$ is measured per-detector with every I--V curve. The amount of blue-leak power $\Pbl$ is estimated from the average optical loading across the ten dark detectors in a module $\Pdd$ .  The average loading across all optical detectors in a module $\Pod$ minus $\Pdd$ provides a per-module estimate of the in-band sky signal power $\Pinband$ coupled through the OMT antennas.

Due to the previously discussed I--V data noise jumps, module 1 and module 2 detector data acquired in dark-lab tests suffer from systematic errors. To avoid unphysical negative optical loading estimates for DD TESs, we apply offsets $\Delta \Psat^{m1} = \SI{1}{\pico\watt}$ and $\Delta \Psat^{m2} = \SI{1.5}{\pico\watt}$ to the module 1 and module 2 optical loading estimates, respectively. These offsets are determined from the average $\Pdd$ measured in the field with the LPF installed. Averaging multiple on-sky DD I--V datasets mitigates systematic errors encountered in the limited set of dark-lab measurements. The optical loading on module 3 and module 4 DD TESs with the LPF installed is $\sim\SI{0.2}{\pico\watt}$. The applied module 1 and module 2 offsets make the average module 1 and module 2 DD optical loading match module 3 and module 4 for the LPF on case. Note that $\Pinband$ is not affected by the applied $\Delta \Psat^{m1,m2}$ offsets, since these equally affect the estimates for $\Popt \approx \Pod+\Delta \Psat^{m1,m2}$ and $\Pbl \approx \Pdd+\Delta \Psat^{m1,m2}$.

\subsubsection{Optical loading results in 2025 with no LPF}
   \begin{figure} 
   \begin{center}
   \begin{tabular}{c} 
   \includegraphics[height=6.5cm]{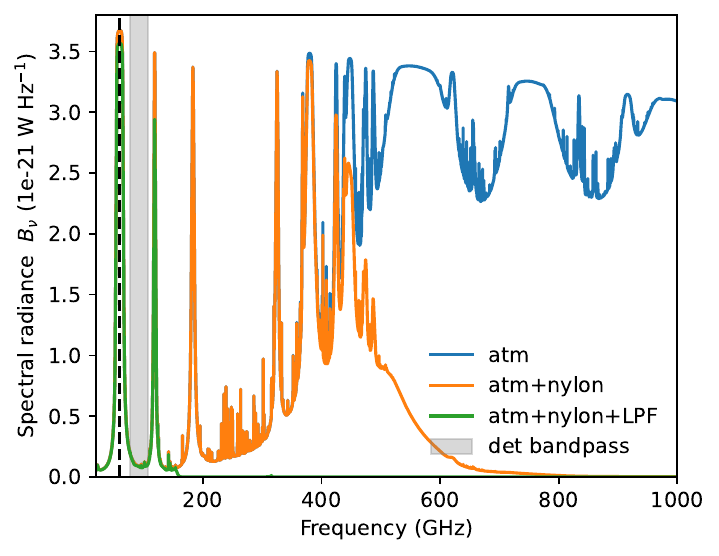}
   \includegraphics[height=6.5cm]{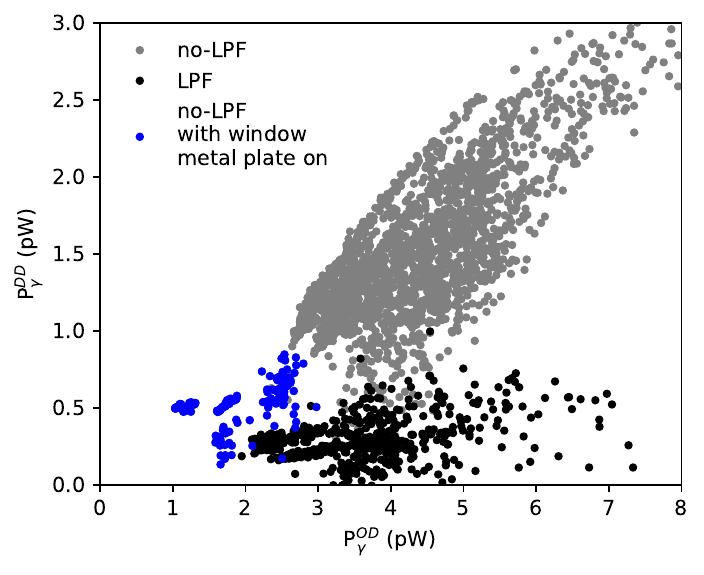}
	\end{tabular}
	\end{center}
   \caption[Atmospheric optical loading and blue leak power] 
   { \label{fig:Pdd} 
\textbf{Left:} In blue, the 50th percentile spectral radiance of the atmosphere from the CLASS site up to \SI{1}{\tera\hertz}\cite{paine24}. In orange, the spectral radiance after multiplying by the \SI{1}{\kelvin} nylon-filter transmission, and in green, after including the LPF. The spectral-radiance calculation presented here assumes single-mode coupling across the entire frequency range and does not include the effects of higher-order waveguide modes or the transmission of the other cryostat IR filters. The dashed vertical line marks the \SI{59}{\giga\hertz} cutoff frequency of the \SI{2.97}{\milli\meter} diameter feedhorn waveguide. The LPF reflects most of the radiation entering through the \SI{1}{\kelvin} nylon filter away from the detector focal plane. \textbf{Right:}  Each data point corresponds to the module-averaged optical-detector loading, $\Pod$, plotted against the module-averaged dark-detector loading, $\Pdd$, for a single I--V curve measurement. The gray data show how $\Pdd$ increases with $\Pod$ for the no-LPF case. This is indicative of blue-leak power coupling directly to both the dark and optical TES. The black points represent data with the LPF installed. The LPF reflects away from the focal plane optical power above \SI{157}{\giga\hertz}, greatly reducing the blue-leak signal. For the LPF data $\Pdd < \SI{0.5}{\pico\watt}$, and  changes little as $\Pod$ and PWV increase, greatly mitigating excess blue-leak NEP. The residual $\Pdd$ power with LPF could be explained by a small temperature offset affecting the focal plane thermometer between on-sky and dark-lab data. The small `islands' of blue data points at low $\Pdd$ and low $\Pod$ correspond to no-LPF data acquired with a metal plate installed in front of the cryostat window.}
   \end{figure}

For the 2025 observing season the CLASS-W2 array observed the sky with no LPF installed at the \SI{50}{\milli\kelvin} stage. The optical loading on module 3 and module 4 OD TESs averaged to \SI{3.4}{\pico\watt}, while for module 1 and module 2 it was \SI{4.6}{\pico\watt} after applying the $\Delta \Psat^{m1,m2}$ offset correction. The DD TESs in each module observe about $\sim\SI{1.1}{\pico\watt}$ of blue-leak power $\Pbl$, which translates to an average module 3 and module 4 in-band power $\Pinband\approx\SI{2.2}{\pico\watt}$, and for module 1 and module 2 $\Pinband\approx\SI{3.3}{\pico\watt}$ (see Table~\ref{tab:tesparam_opt}). The 50\% higher in-band optical loading observed by module 1 and module 2 compared to module 3 and module 4 is a result of the higher optical efficiency of module 1 and module 2, and the higher module 1 and module 2 atmospheric loading due to their average lower elevation angle pointing.

The right plot of Figure~\ref{fig:Pdd} shows the per-module $\Pod$ to $\Pdd$ relationship for on-sky TODs. The gray no-LPF data shows strong positive correlation between $\Pod$ and $\Pdd$ and higher average $\Pdd$ compared to the LPF case. This is indicative of high-frequency power coupling directly to the TES. This blue-leak power measured by $\Pdd$ increases as the atmosphere becomes more opaque with higher precipitable water vapor (PWV), and it was largely suppressed by installing the LPF that reflects optical power above \SI{157}{\giga\hertz} back to the sky.

Note that the 2025 data were mostly acquired during the austral winter when the average PWV at the CLASS site is lower; therefore, we expect the 2025 no-LPF in-band loading to be slightly lower on average than the 2026 LPF data acquired in the austral summer with higher average PWV.

\subsubsection{Optical loading results in 2026 with LPF}
The 2026 LPF dataset includes data acquired between February 2026 and May 2026.
After the CLASS-W2 LPF filter was installed in February 2026, the  optical loading on the dark detectors decreased from $\sim\SI{1.1}{\pico\watt}$ to $\sim\SI{0.2}{\pico\watt}$. This is clear evidence for the blue-leak model, in which the majority of the optical power coupling directly to the TES bolometer island is at frequencies above the \SI{157}{\giga\hertz} cutoff of the LPF (see Figure~\ref{fig:Pdd}). The small remaining \SI{0.2}{\pico\watt} could be explained by one or a combination of the following systematics: 1. A small focal plane thermometer temperature offset between the dark-lab test and the on-sky observations; 2. Optical power below the LPF cutoff coupling directly to the TES; 3. Residual optical power transmitted through the LPF across a wide frequency range. 

With the LPF installed, the optical loading on module 3 and module 4 OD TESs averaged to \SI{2.9}{\pico\watt}, while for module 1 and module 2 it was \SI{4.1}{\pico\watt} after applying the $\Delta \Psat^{m1,m2}$ offset correction, and the DD TESs in each module observe about $\sim\SI{0.2}{\pico\watt}$ of blue-leak power $\Pbl$. Therefore the module 3 and module 4 average in-band power is $\Pinband\approx\SI{2.7}{\pico\watt}$, and for module 1 and module 2 it is $\Pinband\approx\SI{3.9}{\pico\watt}$ (see Table~\ref{tab:tesparam_opt}). The in-band optical loading for the LPF dataset is $\sim\SI{0.6}{\pico\watt}$ higher than for the no-LPF case. This is partly explained by the higher average PWV for the 2026 austral summer of $\SI{1.5}{\milli\meter}$ compared to $\SI{0.8}{\milli\meter}$ for the 2025 dataset. This difference in PWV accounts for $\sim\SI{0.4}{\pico\watt}$ of the in-band loading change between the no-LPF and LPF data. The remaining $\sim\SI{0.2}{\pico\watt}$ could be associated with a higher ambient temperature of the warm optics in the austral summer, or the LPF scattering a small amount of in-band power. Since the LPF is located inside the \SI{1}{\kelvin} volume of the cryostat, mounted on the \SI{50}{\milli\kelvin} stage, we expect optical loading sourced by the LPF emissivity to be negligible. 

The detector’s antenna temperature, $\Ta$, is calculated by dividing the measured in-band optical power, $\Pinband$, by the optical efficiency, $\eta$, and then multiplying by the temperature-to-power conversion coefficient taken from Table~\ref{tab:tesparam_bp}:
 
\begin{equation}
     \Ta  = \frac{\Pinband}{\eta} \frac{d\Trj}{d\Popt} = \frac{\Pinband}{\eta \kb \Delta \nu}.
\label{eqn:Ta}
\end{equation}
Module 4, located at the center of the focal plane, measures an antenna temperature of $\sim\SI{20}{\kelvin}$ (with the LPF installed). The other modules record values about $\sim$10\% higher, due to two effects: 1. extra beam spill onto the warm optics for detectors closer to the focal plane edge; 2. module 1 and module 2 sit at the bottom of the array observing through lower elevation angles that sample a larger atmospheric column and thus a higher effective atmospheric antenna temperature. At the 45 degree elevation angle pointing of the center module, the \SI{90}{\giga\hertz} antenna temperature of the atmosphere together with the CMB account for $\sim\SI{12}{\kelvin}$. The remaining $\sim\SI{8}{\kelvin}$ come from beam spill onto the cold and warm optics, as well as the emissivity of the optical elements. 

When the cryostat window is covered with a metal plate the observed antenna temperature of the module 4 is $\sim\SI{7}{\kelvin}$. Since module 4 beams strike the metal plane at normal incidence, the measured optical loading represents cold optical beam spill plus twice the in-band emissivity of the cryostat’s internal optical elements (including the window) since each pixel sees their direct emission and the reflection from the metal cover. The beam spill onto the cold stage is expected to contribute $\sim\SI{1}{\kelvin}$; therefore, the emissivity of the IR filters, lenses, and window account for $\sim\SI{6}{\kelvin}$ of the loading with the metal plate and therefore $\sim\SI{3}{\kelvin}$ in the on-sky configuration.
Thus, in the on-sky configuration roughly $\SI{4}{\kelvin}$ of the antenna temperature arises from emission by the cryostat optics and cold spill, while an additional $\sim\SI{4}{\kelvin}$ comes from emissivity and beam spill in the warm optical elements—the primary and secondary mirrors, the VPM, and the black forebaffle at the telescope-cage aperture.

\subsection{Noise equivalent power}\label{NEP}

\begin{table}
\caption[NEP]{Median detector NEP across modules. The 1$\sigma$ ranges are computed from the distribution of individual detector values, providing a measure of the uniformity of the parameter across a module. To directly compare the LPF to the no-LPF NEP results, we group measurements and calculate model values at a narrow optical loading range given in the last column of the table by $\Pinband$. As expected from this choice, the photon NEP model ($\nepopt$) contribution are similar for the two datasets. The detector phonon noise $\nepph$  is derived from the parameters in Table~\ref{tab:tesparam_thermal}; therefore it is the same for both cases. $\nepexc^{2} = \nep^{2}-\nepph^{2}-\nepopt^{2}$ includes the remaining contributions from blue-leak noise, readout noise, and any detector noise term not modeled. $\nepexc$ decreases between no-LPF and LPF installed due to the reduction in blue-leak power. Even with the LPF in place we still observe excess noise. This may stem from detector-noise components not captured by our model and/or from residual out-of-band power that continues to reach the TES despite the filter.}% No 1$\sigma$ variance across individual detector values is quoted for $\nepexc$ because it is calculated directly from the other mean NEP values in the table. } 
\label{tab:tesparam_nep}
\begin{center}       
\begin{tabular}{llllcl} 
\hline
\hline
 & \multicolumn{5}{c}{no-LPF} \\
\hline
\rule[-1ex]{0pt}{3.5ex}  Module & $\nep$ (\aWrtS) & $\nepph$ (\aWrtS)& $\nepopt$ (\aWrtS)& $\nepexc$ (\aWrtS)& $\Pinband$ (\SI{}{\pico\watt})\\
\hline
\rule[-1ex]{0pt}{3.5ex}  1 & $35.6\pm2.6$ & $15.9\pm0.6$ & $26.8\pm0.8$ & $17.7\pm4.6$  & $3.7\pm0.1$ \\
\rule[-1ex]{0pt}{3.5ex}  2 & $34.3\pm3.0$ & $14.4\pm0.6$ & $25.5\pm0.9$ & $18.3\pm4.7$  & $3.5\pm0.1$ \\
\rule[-1ex]{0pt}{3.5ex}  3 & $30.0\pm2.4$ & $14.2\pm0.6$ & $19.2\pm0.9$ & $18.1\pm3.6$  & $2.5\pm0.1$  \\
\rule[-1ex]{0pt}{3.5ex}  4 & $27.3\pm2.8$ & $13.9\pm0.6$ & $16.1\pm0.9$ & $16.9\pm4.0$  & $2.0\pm0.1$ \\
\rule[-1ex]{0pt}{3.5ex}  Array & $31.4\pm3.9$ & $14.6\pm0.9$ & $19.9\pm4.0$ & $18.4\pm4.6$  & $2.6\pm0.6$ \\
\hline 
% 17.3, 17.9, 18.3, 17.0, 19.3
%W3 NEP_exc nLPF = 17.7 +- 4.6
%W5 NEP_exc nLPF = 18.3 +- 4.7
%W1 NEP_exc nLPF = 18.1 +- 3.6
%W2 NEP_exc nLPF = 16.9 +- 4.0
%Wall NEP_exc nLPF = 18.4 +- 4.6

 & \multicolumn{5}{c}{LPF} \\
\hline

\rule[-1ex]{0pt}{3.5ex}  1 & $33.5\pm3.5$ & $15.9\pm0.6$ & $26.9\pm0.9$ & $14.2\pm5.5$  & $3.7\pm0.1$ \\
\rule[-1ex]{0pt}{3.5ex}  2 & $32.0\pm3.6$ & $14.4\pm0.6$ & $25.6\pm0.9$ & $14.9\pm6.2$  & $3.5\pm0.1$ \\ 
\rule[-1ex]{0pt}{3.5ex}  3 & $26.9\pm3.9$ & $14.2\pm0.6$ & $19.5\pm0.8$ & $12.2\pm5.7$  & $2.5\pm0.1$  \\
\rule[-1ex]{0pt}{3.5ex}  4 & $24.1\pm4.0$ & $13.9\pm0.5$ & $16.5\pm0.8$ & $11.1\pm6.3$  & $2.0\pm0.1$ \\
\rule[-1ex]{0pt}{3.5ex}  Array & $29.1\pm4.6$ & $14.6\pm0.9$ & $21.2\pm3.9$ & $12.7\pm6.4$  & $2.8\pm0.6$ \\
\hline
% 12.1, 12.5,11.9,10.9,13.6
%W3 NEP_exc LPF = 14.2 +- 5.5
%W5 NEP_exc LPF = 14.9 +- 6.2
%W1 NEP_exc LPF = 12.2 +- 5.7
%W2 NEP_exc LPF = 11.1 +- 6.3
%Wall NEP_exc LPF = 12.7 +- 6.4

\end{tabular}
\end{center}
\end{table}

   \begin{figure} 
   \begin{center}
   \begin{tabular}{c} 
   \includegraphics[height=13cm]{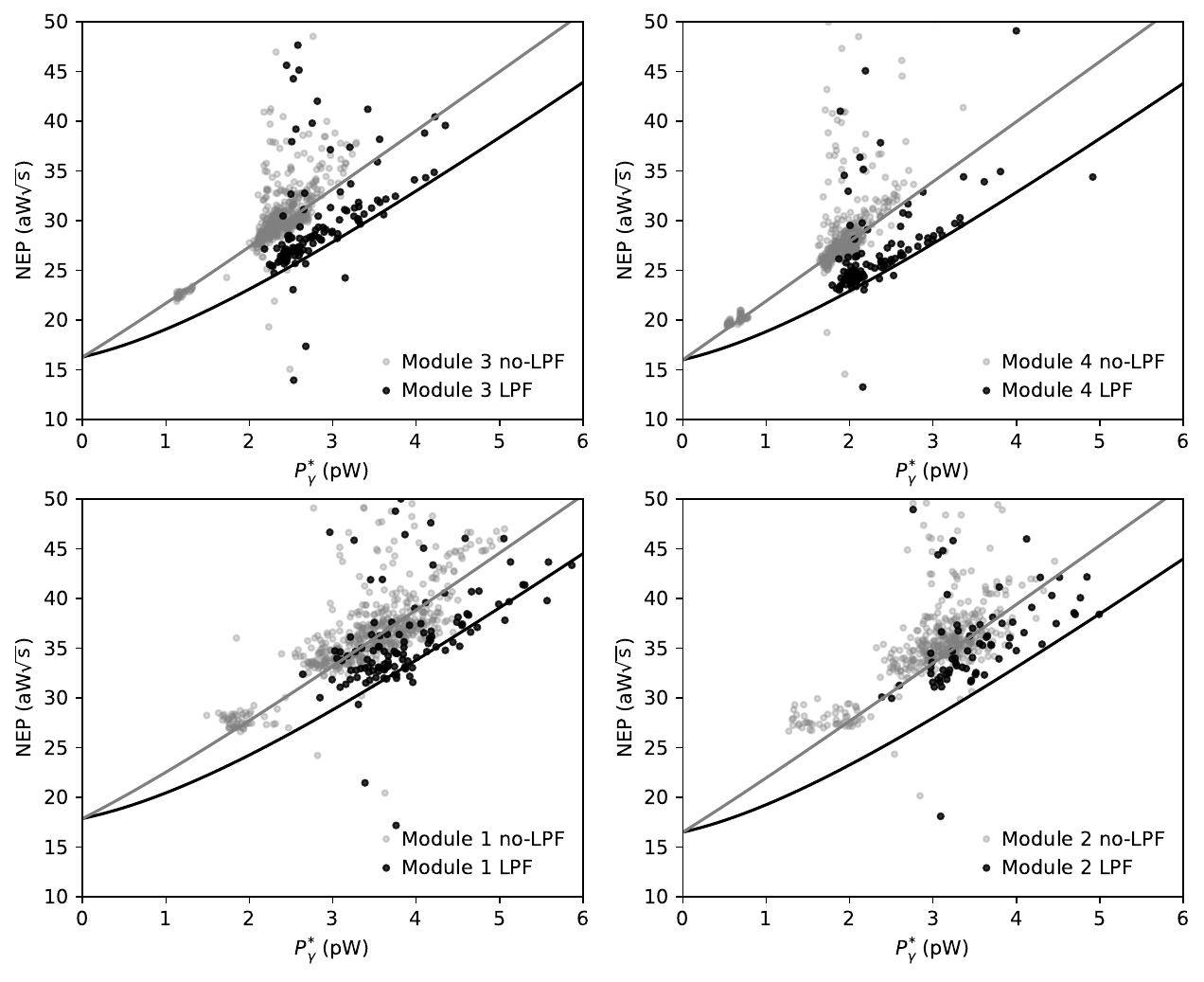}
	\end{tabular}
	\end{center}
   \caption[NEP as a function of optical loading] 
   { \label{fig:NEP_vs_P} 
NEP vs. in-band optical loading ($\Pinband$) plotted per-module, including LPF (black~\textcolor{black}{$\bullet$} dots) and no-LPF (gray~\textcolor{gray}{$\bullet$} dots) data. Each dot corresponds to the median module NEP near the modulation frequency for a 10-minute data package acquired a few minutes after an I--V measurement. The baseline NEP model including detector, readout, and in-band photon noise is marked by the black~{\color{black}\rule[0.5ex]{0.3cm}{2pt}} line. The gray~{\color{gray}\rule[0.5ex]{0.3cm}{2pt}} line marks the NEP baseline model plus a blue-leak contribution that scales with the power observed by the dark detectors $\Pdd$. With the LPF installed the NEP data match the baseline model better, but some excess noise remains. The excess could be associated with additional detector noise not modeled and/or residual out-of-band power.}
   \end{figure}

The noise equivalent power ($\nep$) is calculated by averaging the power spectrum of a TOD near the modulated signal band between \SI{10.5}{\hertz} and \SI{12.5}{\hertz}. Each NEP measurement is associated with a previously acquired I--V curve that sets the target operating bias, produces the responsivity calibration, and provides an estimate of the optical loading during the observations. The NEP measurements in units of $\aWrtS$ are compared with the expected noise model that includes photon, phonon, readout, and blue-leak noise:

\begin{equation}
\nep^{2} = \nepopt^{2}+\nepph^{2}+\nepbl^{2}+\nepsq^{2},
\label{eqn:nep}
\end{equation}

\begin{equation}
\nepopt^{2} =  h \fc \Pinband + \frac{(\Pinband)^{2}}{\Delta \nu}, %\mathrm{and}
\label{eqn:nepopt}
\end{equation}

\begin{equation}
\nepph^{2} = 2 \flink \kb G \Tc^{2},
\label{eqn:nepph}
\end{equation}

\begin{equation}
\nepbl^{2} =  h \fbl \Pbl. 
\label{eqn:nep_bl}
\end{equation}

$\nepsq$ is the SQUID readout noise, with an estimated contribution of $\SI{5}{\atto\watt\sqrt{\second}}$, well below the detector phonon ($\nepph$) and photon noise ($\nepopt$). 
The SQUID readout NEP is estimated by multiplying the measured current readout noise of an isolated SQUID readout channel by the average responsivity of an optical detector. 
The phonon noise ($\nepph$) depends on the thermal conductivity $G$, the critical temperature $\Tc$ and the $\flink$ factor that is approximately $0.5$ since we operate with $\Tb \ll \Tc$\cite{mauskopf18}. Phonon noise is calculated per-module based on the parameters from Section~\ref{thermal_params} and reported in Table~\ref{tab:tesparam_nep}.

Photon noise $\nepopt$ has a shot noise term proportional to the optical loading ($\Pinband$) and the center frequency of the band ($\fc$), and a second bunching noise term is equal to the square of optical power divided by the frequency bandwidth ($\Delta \nu$). Given the CLASS \SI{90}{\giga\hertz} bandpass and on-sky optical loading, the bunching noise term is the dominant contributor. To compare the NEP from similar LPF and no-LPF datasets, we select TODs acquired with an optical loading of \qtylist{ 3.7; 3.5; 2.5; 2.0}{\pico\watt} for modules 1, 2, 3, and 4, respectively. To compare the LPF to no-LPF array average NEP we select data over a wider range of optical loading that includes data from all four modules. These optical loading normalized NEP estimates are used in section~\ref{sec:net} to estimate the array sensitivity. The LPF array average $\nepopt$ is higher due to the higher average PWV during these observations. The average photon noise per-module is reported in Table~\ref{tab:tesparam_nep}. The per-module LPF and no-LPF  $\nepopt$ are similar since they are derived from similar $\Pinband$.    

We include an additional $\nepbl$ term to model noise from out-of-band blue-leak power, which is especially relevant for the 2025 telescope configuration without the LPF. It includes only the shot-noise term, since the blue-leak power coupling directly to the TES is expected to be broadband and centered at a higher frequency than our \SI{90}{\giga\hertz} bandpass. The amount of blue leak power $\Pbl$ is estimated by computing the average optical loading on the dark detectors ($\Pdd$) in a module. In an ideal scenario $\Pdd = 0$, but any excess detected on the DD is expected to affect the OD similarly and contribute to the total NEP. For the no-LPF case $\Pdd\approx\SI{1.1}{\pico\watt}$, assuming a blue-leak center frequency of $\fbl =\SI{400}{\giga\hertz}$ estimated from the transmission of the \SI{1}{\kelvin} nylon filter, yielding $\nepbl\approx\SI{17}{\atto\watt\sqrt{\second}}$. This blue-leak NEP estimate is consistent with the excess NEP reported for the no-LPF case in Table~\ref{tab:tesparam_nep}. 

The per-module measured NEP from on-sky TODs is reported on the second column of Table~\ref{tab:tesparam_nep}. We find it convenient to define excess NEP ($\nepexc$) as:
\begin{equation}
\nepexc^{2} =  \nep^{2}-\nepph^{2}-\nepopt^{2}.
\label{eqn:nep_exc}
\end{equation}
By subtracting the modeled phonon and photon NEP from the measured NEP, we quantify the residual excess noise, which includes contributions from the readout, blue-leak power, and other detector noise sources not included in the model. As shown in Table~\ref{tab:tesparam_nep} $\nepexc$ drops after the LPF was installed, further validating the blue-leak model of direct optical coupling to the TES bolometer island. $\nepexc$ in the LPF case is higher than the expected $\nepbl=\SI{6}{\atto\watt\sqrt{\second}}$ for the measured $\Pdd\approx\SI{0.2}{\pico\watt}$. The main contributor to $\nepexc$ in the LPF case may be intrinsic detector noise, which could be parametrized by a $\flink>0.5$. Figure~\ref{fig:NEP_vs_P} shows the measured NEP plotted versus the measured in-band loading $\Pinband$ for both the LPF and no-LPF datasets. Over-plotted are two NEP models: 1. in black a baseline NEP model including detector, readout, and in-band photon noise that is closer to the LPF data; 2. in gray the baseline NEP model plus a blue-leak contribution that scales with $\Pdd$ and matches the no-LPF data.
The per-detector $\nep$ as a function of detector position in the focal plane is plotted in Figure~\ref{fig:NET_arr}.

\subsection{NET sensitivity}\label{sec:net}

\begin{table}
\caption[NET]{Median detector $\net$ per-module derived from the NEP reported in  Table~\ref{tab:tesparam_nep} for the no-LPF and LPF configurations. The number of working OD TESs is determined by counting the number of detectors with good I--V curve and optical efficiency measurements. The conversion factor $\frac{d\Tcmb}{d\Popt}$ combines the factors in Table~\ref{tab:tesparam_bp} with the optical efficiency reported in Table~\ref{tab:tesparam_opt}. Additionally, we include a $0.98$ LPF optical efficiency factor for data with the LPF installed.} 
\label{tab:tesparam_net}
\begin{center}       
\begin{tabular}{l|lll|lll} 

\hline
\hline
\rule[-1ex]{0pt}{3.5ex}   &   \multicolumn{3}{|c|}{no-LPF} & \multicolumn{3}{c}{LPF} \\
\hline
\rule[-1ex]{0pt}{3.5ex}  Module & $\frac{d\Tcmb}{d\Popt}$ (\SI{}{\kelvin\per\pico\watt})  & $\net$ ($\uKrtS$) & \# of OD & $\frac{d\Tcmb}{d\Popt}$ (\SI{}{\kelvin\per\pico\watt})  & $\net$ (\uKrtS) & \# of OD \\
\hline
\rule[-1ex]{0pt}{3.5ex}  1 & $7.2\pm0.8$ & $260\pm29$ & $70$ & $7.4\pm0.8$  & $254\pm29$ & $70$ \\
\rule[-1ex]{0pt}{3.5ex}  2 & $8.4\pm0.9$ & $292\pm33$ & $67$ & $8.5\pm0.9$  & $282\pm29$ & $67$\\
\rule[-1ex]{0pt}{3.5ex}  3 & $9.6\pm1.0$ & $288\pm28$ & $73$ & $9.8\pm1.1$  & $267\pm27$ & $73$  \\
\rule[-1ex]{0pt}{3.5ex}  4 & $10.6\pm1.5$ & $284\pm30$ & $67$ & $10.8\pm1.6$  & $266\pm27$ & $68$ \\
\rule[-1ex]{0pt}{3.5ex}  Array & $8.9\pm1.7$ & $278\pm32$ & $277$ & $9.1\pm1.7$  & $268\pm31$ & $278$ \\
\hline 
\end{tabular}
\end{center}
\end{table}

The noise equivalent temperature ($\net$) combines the measured $\nep$ with the optical efficiency and bandpass measurements to yield a per detector sensitivity estimate:
\begin{equation}
\net = \frac{1}{\eta}\frac{d\Tcmb}{d\Trj} \frac{d\Trj}{d\Popt} \nep
\label{eqn:net}
\end{equation}
$\net$ is proportional to the ratio of $\nep$ noise to optical efficiency $\eta$ calibrated to CMB temperature units by converting power at the detector to $\Trj$ with $\frac{d\Trj}{d\Popt}$ and then to CMB temperature units with $\frac{d\Tcmb}{d\Trj}$. 
The median detector $\net$ sensitivity is computed per module and reported in Table~\ref{tab:tesparam_net}, along with the number of working optical detectors. For a polarization sensitive experiment such as CLASS, $\net$ is converted to linear polarization sensitivity $\nequ$ by dividing $\net$ by the polarization modulation efficiency. The VPM, which modulated between circular and linear polarization, has a linear polarization modulation efficiency of $\sim0.7$, while the RHWP, which modulates between $Q$ and $U$ Stokes parameters, has a linear modulation efficiency of $\sim1.0$.

\begin{figure} 
   \begin{center}
   \begin{tabular}{c} 
   \includegraphics[height=6.5cm]{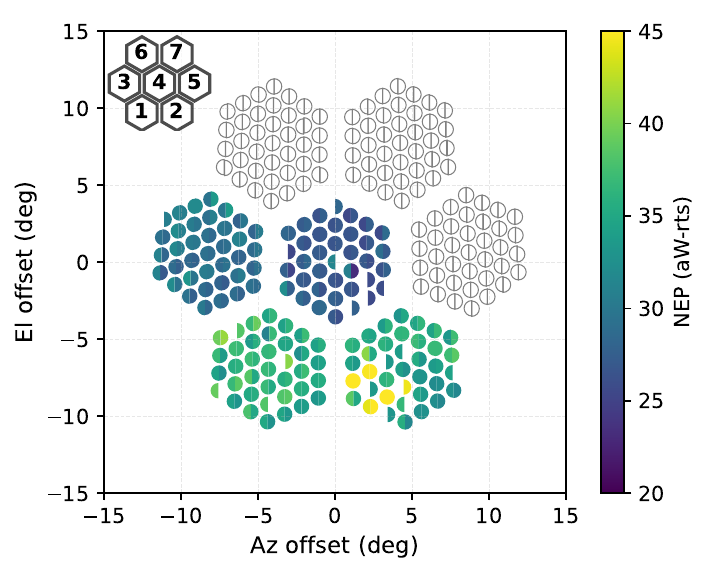}
   \includegraphics[height=6.5cm]{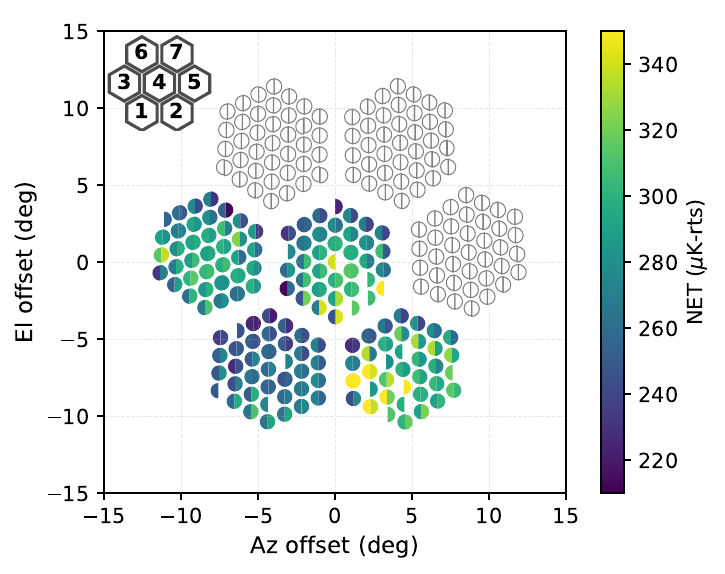}
	\end{tabular}
	\end{center}
   \caption[NET and NEP array plots] 
   { \label{fig:NET_arr} 
\textbf{Left:} CLASS-W2 array plot of NEP. \textbf{Right:} CLASS-W2 array plot of NET. Each detector’s sensitivity is displayed at its corresponding position on the focal plane. }
   \end{figure}

The per-detector $\net$ as a function of detector position in the focal plane is plotted in Figure~\ref{fig:NET_arr}. The median per-detector NET across the array improved after the LPF installation to $\net = 268~\uKrtS$. Each module has around 70 working OD TESs, and a sensitivity of $\net = 32~\uKrtS$. We identify working OD TESs by counting those that display both a reliable I–V curve and a valid optical efficiency measurement. The partial four module CLASS-W2 focal plane has $\net = 16~\uKrtS$, which enhances the CLASS \SI{90}{\giga\hertz} mapping speed by 41\%\cite{nunez25}. Roughly 4\% of the improvement in mapping speed is attributable to the faster detector time constants in CLASS-W2 relative to CLASS-W1. Three new detector modules are under fabrication to complete the seven module CLASS-W2 focal plane. The new detectors target lower $\Tc$, which will improve per-detector sensitivity and further enhance the mapping speed of CLASS-W2. The deployment of the new modules is scheduled for 2027.

\section{conclusions}\label{sec:conclusions}

We have deployed and characterized, through dark-lab and on-sky observations, the performance of the first four modules of the CLASS-W2 detector focal plane. This version of the CLASS-W2 $\SI{90}{\giga\hertz}$ telescope achieves an array sensitivity of $\net = 16~\uKrtS$ and improves the CLASS CMB mapping speed by $41\%$, with further improvement expected in 2027 when three additional modules will be installed.

On-sky observation with and without a free-space metal-mesh LPF installed in front of the focal plane shows that high frequency out-of-band `blue-leak' optical power couples directly to the TES bolometer island affecting both dark (DD) and optical detectors (OD). By installing the LPF we reduced the `blue leak' power by $\SI{0.9}{\pico\watt}$  and the per-detector NEP$^{2}$ by $(\SI{13}{\atto\watt\sqrt{\second}})^{2}$. After accounting for the LPF transmission, the array sensitivity improves by 4\%. 
Even with the LPF installed, we still measure excess noise comparable to the detector phonon noise.  This excess may be explained by a combination of noise contributions including: readout noise, higher $\flink$ factor in the phonon noise model, additional detector noise sources not included in the model, and residual out-of-band power below the LPF cutoff coupling directly to the TES.
The CLASS-W2 average in-band detector optical loading of $\SI{3.3}{\pico\watt}$ is consistent with the loading seen by the CLASS-W1 telescope when differences in optical efficiencies are taken into account. This loading translates to an average focal plane antenna temperature of \SI{23}{\kelvin} (\SI{20}{\kelvin} for the center module that is subject to less warm spill), with a \SI{13}{\kelvin} contribution from the atmosphere and the CMB, and a \SI{10}{\kelvin} contribution from the optical spill and emissivity of the cold and warm optical components. 

The end-to-end optical efficiency including detector and telescope optics for module 3 and module 4 is about 0.33, while for module 1 and module 2 it is higher at 0.42. The module 3 and module 4 optical efficiency is about half the efficiency of the CLASS-W1 telescope upgraded detector modules. To maximize optical efficiency, the CLASS-W1 detectors employ monocrystalline silicon as the dielectric in their microwave components, reducing loss and enabling precise bandpass fabrication. On the other hand the detector yield of the CLASS-W1 telescope is 66\%\cite{nunez25} compared to the 94\% yield achieved by the CLASS-W2 detectors. Additionally, the $\Tc=\pm\SI{3}{\milli\kelvin}$ uniformity of the CLASS-W2 AlMn TES is exceptionally good, compared to the larger $\Tc=\pm\SI{8.5}{\milli\kelvin}$ for the CLASS-W1 MoAu TES.
Fine-tuning the AlMn recipe\cite{duff26,dutcher26} to achieve the average targeted $\Tc = \SI{160}{ \milli\kelvin} $ is a principal goal of the ongoing detector fabrication of the remaining three CLASS-W2 modules. 

The achieved array averaged detector bandpass has a center frequency of \SI{95.2}{\giga\hertz}, \SI{2.6}{\giga\hertz} higher than the simulated bandpass. The \SI{28.3}{\giga\hertz} bandwidth is \SI{2.7}{\giga\hertz} below the target. Increasing the detector bandwidth in future fabrication cycles is one path to improve detector sensitivity. Nevertheless, the realized bandpass falls well within the atmospheric band, allowing for high sensitivity mapping of the CMB. 
Observations of Jupiter provide a measure of the total optical efficiency and detector beam profile. The array average beam solid angle $\Omega = \SI{124\pm0.3}{\micro\steradian}$, and Full-Width-Half-Maximum angle of $\SI{0.592\pm0.003}{\deg}$  match expectations. Moon observations reveal side-lobe pick up at $<0.1\%$ level that is co-aligned with other feedhorns in the same module. The amplitudes of these optical cross-talk signals are suppressed after installing the LPF.

The responsivity of the detectors is estimated from on-sky I--V data. For module 3 and module 4 the I--V slope responsivity closely matches $S^*$ when biased low on the transition,  as expected in the high $\Lg$ limit. For module 1 and module 2 the I--V slope responsivity is $\sim$20\% larger than $(S^*)^{-1}$. This unexpected difference is under investigation but was confirmed by multi-$\Tb$ responsivity measurements. 
The expected TES thermal time constant is $\tau = \SI{6.5}{\milli\second}$, while the measured optical time constants, sped-up by electro-thermal feedback, are about ten times faster. This speed-up factor is consistent with the average array $\Lg = 24$ and an inferred $\beta = 1.5$. 

Multi-$\Tb$ I--V data are combined across module 3 and module 4 detectors to determine the TES $\alpha$ parameter as a function of TES resistance. The detectors are typically biased near the middle of the transition, at half their normal resistance; therefore we expect $\alpha\approx100$ during typical CMB observations.
From dark-lab multi-$\Tb$ I--V data we also extract the average TES thermal parameters: $\Tc = \SI{184}{ \milli\kelvin}$, $\kappa =\SI{10.4}{\nano\watt\per\kelvin^n}$, $G = \SI[per-mode = symbol]{460}{\pico\watt\per\kelvin}$, $\Pph = \SI{23.3}{\pico\watt}$, $n = 3.6$,  and $\Rn = \SI{7.8}{\milli\ohm}$.
The average achieved $\Tc$ is $\SI{24}{ \milli\kelvin}$ above the target  $\Tc = \SI{160}{ \milli\kelvin}$. This means the detectors have higher saturation power than initially targeted but also higher detector NEP. The high saturation power is convenient for observing bright sources like the Moon, but to optimize sensitivity we prefer to reduce $\Tc$ and cut the detector saturation power by half.

\appendix    %>>>> this command starts appendixes

% Reset the figure counter and prefix it with the appendix letter
\counterwithin{figure}{section} 
\counterwithin{table}{section}

\section{CLASS detector focal plane deployment history}

Over the lifetime of the CLASS project, we have deployed new and upgraded detector focal planes operating at \qtylist[list-units = single]{40; 90; 150; 220}{\giga\hertz}. Table~\ref{tab:hist_fp} summarizes the deployment dates and key characteristics of these new or upgraded focal planes. The baseline CLASS telescope configuration we have followed so far includes two mounts, the first mount points the CLASS-Q (\SI{40}{\giga\hertz}), and CLASS-W1 (\SI{90}{\giga\hertz}) telescopes, while mount 2 points the CLASS-HF (\SI{150,220}{\giga\hertz}) and CLASS-W2 (\SI{90}{\giga\hertz}) telescopes.

\begin{table}
\caption[CLASS upgrade history]{CLASS detector focal plane deployment and upgrade history. } 
\label{tab:hist_fp}
\begin{center}       
\begin{tabular}{lllll} 
\hline
\hline
\rule[-1ex]{0pt}{3.5ex}  \textbf{Date} &  \textbf{Telescope} & \textbf{Type} & \textbf{Foundry} & \textbf{Description} \\
\rule[-1ex]{0pt}{3.5ex}  June 2016 & CLASS-Q & new & NASA GSFC & 72 detectors\cite{appel19}  \\
\rule[-1ex]{0pt}{3.5ex}  May 2018 & CLASS-Q & upgrade & NASA GSFC & Recovered readout row\cite{dahal22}  \\
\rule[-1ex]{0pt}{3.5ex}  May 2018 & CLASS-W1 & new & NASA GSFC & 7 modules, 518 detectors\cite{dahal18spie}\cite{dahal22}  \\
\rule[-1ex]{0pt}{3.5ex}  September 2019 & CLASS-HF & new & NASA GSFC & 3 dichroic modules, 1020 detectors\cite{dahal20HF}\cite{dahal22}  \\
\rule[-1ex]{0pt}{3.5ex}  August 2022 & CLASS-W1 & upgrade & NASA GSFC & \makecell[l]{Replaced 4 modules with upgraded \\ high optical efficiency detectors\cite{nunez22spie}\cite{nunez25}}   \\
\rule[-1ex]{0pt}{3.5ex}  August 2025 & CLASS-W2 & new & NIST Boulder & 4 modules, 296 detector (this work)  \\
\rule[-1ex]{0pt}{3.5ex}  March 2026 & CLASS-W2 & upgrade & NIST Boulder & Installed LPF (this work)  \\
\hline
\hline
\end{tabular}
\end{center}
\end{table}

\section{Detector wafer vertical profile and dimensions}
   \begin{figure} 
   \begin{center}
   \begin{tabular}{c} 
   \includegraphics[height=8cm, trim={3cm 4cm 2cm 4cm},clip]{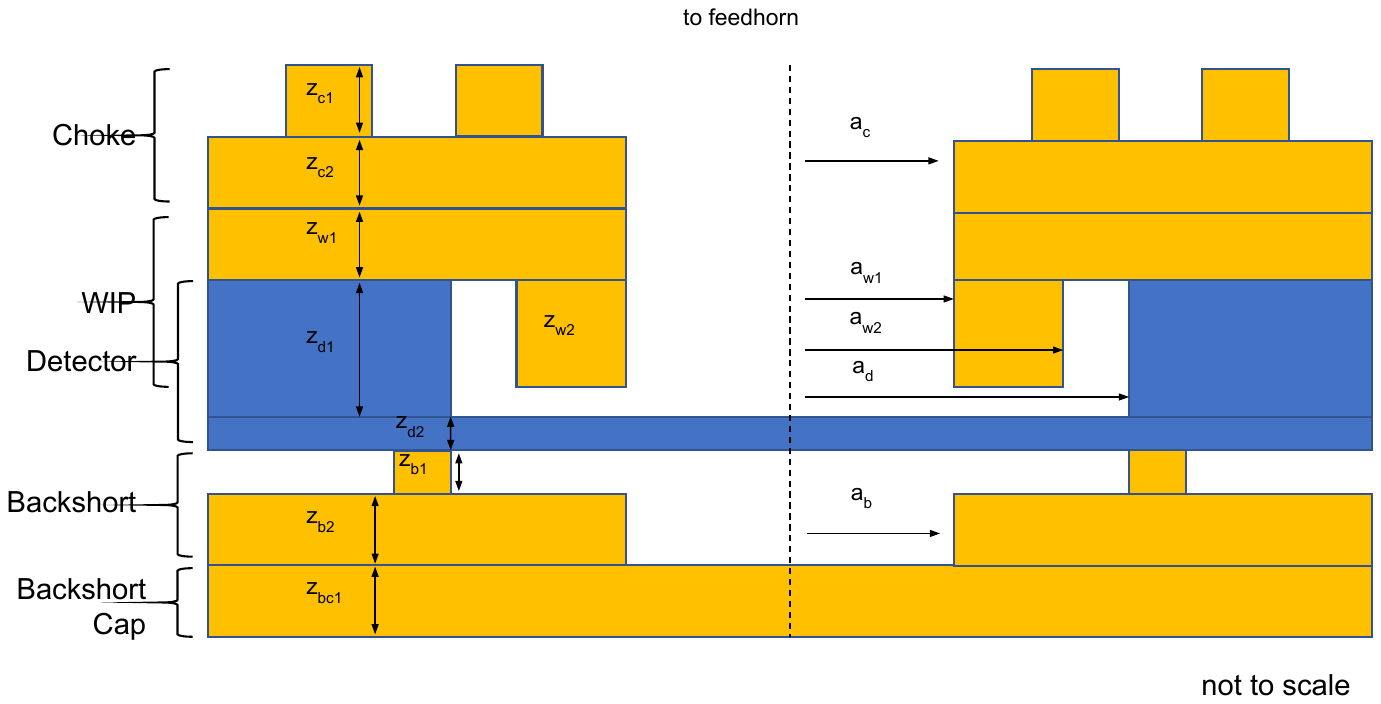}
	\end{tabular}
	\end{center}
   \caption[Detector silicon stack] 
   {\label{fig:det_stack} 
   Detector silicon stack (diagram not to scale). The stack contains four main components: 1. the choke wafer, 2. the waveguide interface plate (WIP), 3. the detector wafer, and 4. the backshort. The components colored in orange are gold plated. The dimensions of features labeled in the figure are summarized in Table~\ref{tab:stack_dim}. }
   \end{figure} 
 This appendix describes the vertical profile of an integrated detector wafer including Choke, WIP, Detector, Backshort, and Backshort Cap.  Figure~\ref{fig:det_stack} shows a diagram of the vertical stack of detector wafers, with the dimensions summarized in Table~\ref{tab:stack_dim}.

\begin{table}
\caption[Dimension of the detector silicon stack]{Dimension of the detector silicon stack shown in Figure~\ref{fig:det_stack}} 
\label{tab:stack_dim}
\begin{center}       
\begin{tabular}{llll} 
\hline
\hline
\rule[-1ex]{0pt}{3.5ex}  \textbf{Part} &  \textbf{Parameter} & $\SI{}{\bf\SI{}{\micro\meter}}$ & \textbf{Description} \\
\hline
\rule[-1ex]{0pt}{3.5ex}  Choke & Z$_\mathrm{c1}$ & 100 & choke pillar height  \\
\rule[-1ex]{0pt}{3.5ex}  Choke & Z$_\mathrm{c2}$ & 500 & choke base thickness  \\
\rule[-1ex]{0pt}{3.5ex}  Choke & a$_\mathrm{c}$ & 1485 & waveguide radius  \\
\hline
\rule[-1ex]{0pt}{3.5ex}  WIP & Z$_\mathrm{w1}$ & 500 & WIP base thickness  \\
\rule[-1ex]{0pt}{3.5ex}  WIP & Z$_\mathrm{w2}$ & 485 & WIP re-entrant depth  \\
\rule[-1ex]{0pt}{3.5ex}  WIP & a$_\mathrm{w1}$ & 1485 & boss ID / waveguide radius  \\
\rule[-1ex]{0pt}{3.5ex}  WIP & a$_\mathrm{w2}$ & 2285 & boss OD   \\
\hline
\rule[-1ex]{0pt}{3.5ex}  Detector & Z$_\mathrm{d1}$ & 500 & wafer thickness \\
\rule[-1ex]{0pt}{3.5ex}  Detector & Z$_\mathrm{d2}$ & 2.45 & thickness of SiN+SiOx \\
\rule[-1ex]{0pt}{3.5ex}  Detector & a$_\mathrm{d}$ & 2335 &   OMT probe membrane radius \\
\hline
\rule[-1ex]{0pt}{3.5ex}  Backshort & Z$_\mathrm{b1}$ & 10 & Si standoff to OMT probe plane \\
\rule[-1ex]{0pt}{3.5ex}  Backshort & Z$_\mathrm{b2}$ & 665 & backshort depth - standoff\\
\rule[-1ex]{0pt}{3.5ex}  Backshort & a$_\mathrm{b}$ & 1485 &  waveguide radius\\
\rule[-1ex]{0pt}{3.5ex}  Backshort cap & Z$_\mathrm{bc1}$ & 500 &  backshort cap thickness\\

\hline
\hline
\end{tabular}
\end{center}
\end{table}

%\acknowledgments % equivalent to \section*{ACKNOWLEDGMENTS}       
\begin{acknowledgments}

We acknowledge the National Science Foundation Division of Astronomical Sciences for their support of CLASS under Grant Numbers 0959349, 1429236, 1636634, 1654494, 2034400, 2109311, and 2442928.
We thank Johns Hopkins University President R. Daniels and the Krieger School of Arts and Sciences Deans for their steadfast support of CLASS.
We further acknowledge the very generous support of Jim Murren and Heather Miller (JHU A\&S '88), Matthew Polk (JHU A\&S Physics BS '71), David Nicholson, and Michael Bloomberg (JHU Engineering '64).
The CLASS project employs detector technology developed in collaboration between JHU and Goddard Space Flight Center under several NASA grants. Detector development work at JHU was funded by NASA cooperative agreement 80NSSC19M0005. 
J. Eimer acknowledges support from NASA award number 80NSSC25M7076.
%We acknowledge scientific and engineering contributions from Max Abitbol, Aamir Ali, Fletcher Boone, Ricardo Bustos, David Carcamo, Manwei Chan, David Chuss, Lance Corbett, Rahul Datta, Rolando Dünner Planella, Joseph Golec, Dominik Gothe, Ted Grunberg, Mark Halpern, Saianeesh Haridas, Kathleen Harrington, Jake Hassan, Connor Henley, Kyle Helson, Jeff Iuliano, Ben Keller, Lindsay Lowry, Jeffrey J. McMahon, Nick Mehrle, Nathan Miller, Sasha Novak, Carolina Nunez, Keisuke Osumi, Ivan Padilla, Gonzalo Palma, Lucas Parker, Diva Parekh, Isu Ravi, Rodrigo Reeves, Gary Rhodes, Carl Reintsema, Karwan Rostem, Daniel Swartz, Bingjie Wang, Qinan Wang, Duncan Watts, Tiffany Wei, Janet Weiland, Zhilei Xu, Zi´ang Yan, Lingzhen Zeng, and Zhuo Zhang.
We thank Miguel Angel D\'iaz, Heather Skrzecz, and Sherrell Byrd-Arthur for logistical support. We acknowledge productive collaboration with the JHU Physical Sciences Machine Shop team.
CLASS is located in the Parque Astron\'omico Atacama in northern Chile under the auspices of the Agencia Nacional de Investigaci\'on y Desarrollo (ANID). 
We acknowledge the assistance of HopGPT, which was used to refine grammar and style; all scientific content remains the sole responsibility of the authors.
This manuscript is currently under consideration for publication as an SPIE Astronomical Telescopes + Instrumentation 2026 conference proceedings paper, Paper Number 14156-55.

\paragraph{Software:}\textit{numPy}~\citenum{numpy20}, \textit{sciPy}~\citenum{scipy},
\textit{matplotlib}~\citenum{matplotlib}, \textit{pyephem} \citenum{pyephem}, and \textit{getdata} \citenum{getdata}.
% pyephem \citep{pyephem}, }
%\textit{astropy}~\citenum{astropy}
%camb \citep{camb},
%fastcc \citep{2022RNAAS...6..252P},
%pysm \citep{pysm},
%PolSpice \citep{polspice},
%NaMaster \citep{Alonso:2018jzx}
% emcee \citep{emcee},\textit{NumPy} \citep{harris2020numpy}, \textit{SciPy} \citep{virtanen2020scipy}, and \textit{Astropy} \citep{astropy2018}.

\end{acknowledgments}

% References
\bibliography{class_pub,cmb,cosmology,hardware,software} % bibliography data in report.bib
\bibliographystyle{spiebib} % makes bibtex use spiebib.bst

%%%%% Biographies of authors %%%%%

\vspace{2ex}\noindent\textbf{John William Appel}  is an associate research scientist at Johns Hopkins University whose work focuses on cryogenic detector instrumentation for measuring the cosmic microwave background (CMB). He received dual B.S. degrees in physics and mathematics from the University of Miami in 2006 and earned his Ph.D. in physics from Princeton University in 2012, where he developed detectors for the Atacama B-mode Search. At the Cosmology Large Angular Scale Surveyor (CLASS), Dr. Appel led the design, assembly, and on-sky validation of multi-frequency TES-bolometer arrays (40–220 GHz), which produced the first ground-based, large-scale CMB‐polarization maps used to constrain the optical depth to reionization.  He is the author or co-author of more than 50 refereed journal papers and is a member of SPIE.

\listoffigures
\listoftables

\end{spacing}
\end{document}